\documentclass[letterpaper,showpacs,aps]{revtex4-2}
\usepackage[utf8]{inputenc}
\usepackage[spanish]{babel}
\usepackage{latexsym}
\usepackage{amssymb}
\usepackage{amsmath}
\usepackage{bm} 
\usepackage[margin=100pt]{geometry}

\usepackage{epsfig}
\usepackage{graphicx}
\usepackage{subcaption}
\usepackage{float}
\usepackage{color}

\usepackage{algpseudocode}
\usepackage{algorithm}
\usepackage{algorithmicx}
\algdef{SE}[DOWHILE]{Do}{doWhile}{\algorithmicdo}[1]{\algorithmicwhile\ #1}%
\usepackage{listings}
\usepackage{color}
\definecolor{dkgreen}{rgb}{0,0.6,0}
\definecolor{gray}{rgb}{0.5,0.5,0.5}
\definecolor{mauve}{rgb}{0.58,0,0.82}
\definecolor{backcolour}{rgb}{0.95,0.95,0.92}
\newtheorem{theorem}{Teorema}[section]

\newtheorem{definition}{Definición}[section]
\numberwithin{equation}{section}
\numberwithin{equation}{subsection}

\begin{document}


\title{Algoritmo Concurrente por Conjuntos de Pilas con Multiplicidad: SetStackLogic}

\author{José Damián López}
\email{damian13.03@ciencias.unam.mx}


\affiliation{Universidad Nacional
Aut\'onoma de M\'exico, Ciudad de M\'exico 04510, M\'exico.}


\date{\today}


\begin{abstract}
El presente artículo tiene como objetivo describir y explicar los fundamentos teóricos de algoritmo concurrente y concurrentes por conjuntos, considerando un sistema de memoria compartida asincrónica donde cualquier número de procesos puede colapsar. La verificación de los algoritmos concurrentes se describe a menudo a partir de su condición de progreso, la cual garantiza que eventualmente algo bueno sucederá, también llamada la seguridad de los algoritmos, y la correctitud, la cual garantiza que nada malo sucederá, también llamada viveza de los algoritmos. Se explica a detalle el significado de corrección de un algoritmo concurrente, centrándonos en la linealizabilidad, y se aborda una generalización, la concurrencia por conjuntos; la cual es mucho más reciente y menos conocida. Se muestra el algoritmo {\it SetStackLogic}, el cual es un algoritmo concurrente por conjuntos y además es una implementación de una pila con multiplicidad. Se demuestran, de una manera formal y detallada, las propiedades del algoritmo {\it SetStackLogic}, esto con el fin de presentar un esquema riguroso en la formalización de este tipo de algoritmo; mismo que podría ser utilizado para otros algoritmos. Además, se explica el funcionamiento del algoritmo mediante ejemplos de escenarios que ilustran su dinámica en algunas posibles ejecuciones.
\end{abstract}


\maketitle


\section{\label{sectionI}Introducción}

\subsection{Contexto}

El estudio de la teoría de la computación determina si un problema es o no computable; es decir, ser resuelto por medio de un algoritmo, así como su eficiencia. El estudio de los algoritmos comenzó con Turing, cuyos resultados llevaron a formalizar el concepto de algoritmo y de máquina de Turing, lo cual llevo hasta la {\it tesis de Church-Turing}. Aún más, las Máquinas de Turing pueden ser descritas por medio de modelos multicinta y estas son equivalentes, en su poder computacional, a las máquinas secuenciales. Dicho en otras palabras, ambos modelos son equivalentes en cuanto a los problemas que pueden resolver. Sin embargo, eventualmente fue necesario desarrollar un concepto formal sobre la eficiencia de los algoritmos, la {\it complejidad}. Con lo cual se busca desarrollar y estudiar algoritmos que puedan resolver los problemas en un tiempo razonable, y si bien diferentes modelos de cómputo sean equivalentes, estos pueden llegar poseen un desempeño considerablemente diferente. Una de las soluciones, para disminuir la complejidad, han sido los algoritmos que puedan ejecutar procesos de forma simultánea.\\

Por otro lado, las arquitecturas multinúcleo han experimentado un gran auge, pues este diseño evita el sobrecalentamiento generado en un solo núcleo o procesador que trabaja a altas velocidades. Estas arquitecturas multinúcleo son descritas por medio de modelos de cómputo concurrente \cite{ArtOfMpP}. Lo cual permite que, para aumentar la eficiencia, el objetivo cambie a explotar el paralelismo, siendo este uno de los desafíos más destacados en cómputo moderno.\\

\subsection{La Relajación de Multiplicidad}

Recientemente, se ha buscado mejorar el rendimiento de objetos concurrente relajando su semántica. En particular, varios estudios se han centrado en la relajación de colas y pilas, logrando mejoras significativas en el rendimiento \cite{DistributedQueuesSharedMemoryMulticorePerformanceScalabilityQuantitativeRelaxation,QuantitativeRelaxationConcurrentDataStructures,PerformanceScalabilitySemanticsConcurrentFIFOQueues}. Una relajación de estas estructuras es la multiplicidad \cite{Castañeda2022, FullyReadWrite,RelaxedQueuesAndStacks}. Intuitivamente, en una estructura con multiplicidad ciertos conjuntos de operaciones simultáneas actúan de modo que el {\it estado} del objeto cambie como si solo una sola operación hubiera  sido ejecutada. Dicho de otro modo, múltiples operaciones actúan como una misma \cite{michael2009idempotent}.  Estos objetos concurrentes, como contadores, colas, pilas, conjuntos y otros objetos, son implementados en sistemas asincrónicos, donde los procesos simultáneos se comunican accediendo a una memoria compartida y estos son propensos a fallas. 

\subsection{Organización}
En la sección~\ref{sectionII} se dan los fundamentos teóricos así como un modelo de computación concurrente, se describen las nociones formales de operaciones y concurrencia de operaciones. En la sección~\ref{sectionIII} se explica y se describe formalmente la noción de correctitud para un algoritmo concurrente. En la sección~\ref{sectionIV} se muestra el algoritmo {\it SetStackLogic}, el cual es un algoritmo concurrente por conjuntos que es una implementación de pila con multiplicidad; en la literatura se pueden encontrar diversos objetos concurrentes, así como otro tipo de semánticas relajadas, como objetos concurrentes por intervalos. En la sección~\ref{sectionV} se da una demostración formal y rigurosa sobre las propiedades del algoritmo {\it SetStackLogic}, la cual tiene como propósito explicar el enfoque formal de una manera clara y precisa, mismo que puede ser empleado en otros objetos concurrentes. Por otro lado, en la sección~\ref{sectionVI} se explica, por medio de ejemplos, el funcionamiento del algoritmo {\it SetStackLogic}.

\section{\label{sectionII} Fundamentos}
En un algoritmo concurrente, varios hilos se coordinan para modificar de manera simultánea la información que se encuentran almacenados en una memoria compartida. Uno de los retos más difíciles de solucionar en este tipo de algoritmos es el de coordinar las modificaciones concurrentes de forma tal que los datos en la memoria compartida siempre estén en un estado consistente, de forma que no ocurran fallas inesperadas que no se llegan a encontrar en el cómputo secuencial. \\

Formalmente, se puede entender un proceso como la transición de estados de una máquina de estado, donde estas transiciones de estado las denominamos eventos. Se le denota a las transiciones de estados como:
\begin{equation*}
    \delta (\boldsymbol{\hat{q}},\boldsymbol{e}  ) = \boldsymbol{\hat{p}},
\end{equation*}
donde $\boldsymbol{e}$ es un evento, $\boldsymbol{\hat{q}}$ es el estado del proceso {\it antes} del evento, $\boldsymbol{e}$ y  $\boldsymbol{\hat{p}}$ el estado del proceso {\it después} del evento $\boldsymbol{e}$. Esto tiene una interpretación clara, un proceso concurrente es una máquina de estado y los eventos son transiciones de estado (ejecuciones de instrucciones, las cuales podrían dejar invariante el estado de la máquina).
\\

Los eventos son instantáneos: ocurren en un solo instante de tiempo. Es conveniente exigir que los eventos nunca sean simultáneos, es decir, eventos distintos ocurren en momentos distintos, esto nos permitirá inducir un orden total en el conjunto de eventos de la ejecución de los algoritmos concurrentes. \\

Un hilo $A$, produce una secuencia de eventos $a_0,a_1...a_n$. Vamos a denotar la $j-$esíma ocurrencia de un evento $a_i$ como $a_i^{j}$. Se dice que un evento $a$ precede otro evento $b$, cuando, $a$ ocurre antes que $b$, lo denotamos como $a\rightarrow b$. La relación de precedencia $\rightarrow$ es un orden total de eventos. También podemos dar una noción del tiempo que ocurre entre dos eventos. Sean $a_0,a_1$ eventos tal que $a_0\rightarrow a_1$, se define el intervalo $I(a_0,a_1)$, como la duración entre $a_0, a_1$. Se dice que un intervalo  $I_A(a_0,a_1)$ precede a otro intervalo $I_B(b_0,b_1)$, cuando $a_1\rightarrow b_0$, lo denotamos como $I_A\rightarrow I_B$. Esta relación es un orden parcial en los intervalos. Se definen como intervalos concurrentes a aquellos que no están relacionados, es decir si $\neg(I_A\rightarrow I_B \wedge I_B\rightarrow I_A)$.\\

Estas definiciones permiten describir, de manera clara, las operaciones y/o métodos de cualquier algoritmo, incluso describir cuando estas son concurrentes. Aún más, bajo este esquema una operación es un intervalo, donde la invocación o llamada a la operación es un evento, y posteriormente la finalización de dicha operación o retorno es el evento que marca el fin de la operación. 

\subsection{Modelo de Computación}

Vamos a considerar el modelo estándar de sistemas concurrentes, el cual ha sido usado en \cite{RelaxedQueuesAndStacks,FullyReadWrite}, con $n$ procesos asíncronos, $p_1, . . . , p_n$ los cuales, pueden llegar a fallar durante una ejecución; formalmente un proceso falla cuando deja de dar pasos. El {\it índice} del proceso $p_i$ es $i$. Los procesos se comunican entre sí invocando operaciones atómicas en objetos base compartidos. Un {\it objeto base} puede proporcionar operaciones atómicas de lectura/escritura, a partir de ahora, dicho objeto se denomina {\it registro}, u operaciones atómicas más potentes de lectura, modificación y escritura.\\

Un {\it objeto concurrente}, $T$, se define como una máquina de estado que consta de: un conjunto de estados, un conjunto finito de operaciones y un conjunto de transiciones entre estados. Esta especificación no necesariamente tiene que ser secuencial, es decir:
\begin{itemize}
\item  Un estado puede tener operaciones pendientes
\item Las transiciones de estado pueden involucrar varias invocaciones
\end{itemize}
La noción de objeto concurrente se formaliza en las siguientes subsecciones.\\

Una implementación de un objeto concurrente $T$ es un algoritmo distribuido $A$ que consta de máquinas de estados locales $A_1, . . . ,A_n$. Cada máquina local $A_i$ especifica qué operaciones en los objetos base, $p_i$ ejecuta para devolver una respuesta cuando invoca una operación de alto nivel de $T$. Cada una de estas invocaciones de operación en los objetos base es un paso.\\

Una {\it ejecución} de $A$ es una secuencia (posiblemente infinita) de pasos, es decir, ejecuciones de operaciones de objetos base, más invocaciones y respuestas a operaciones del objeto concurrente $T$, con las siguientes propiedades:
\begin{enumerate}
\item Cada proceso es secuencial. Primero invoca una operación, y solo cuando tiene su respuesta correspondiente, puede invocar otra operación, es decir, las ejecuciones están bien formadas.
\item Para cualquier invocación a una operación $op$, denotada $inv(op)$, de un proceso $p_i$, los pasos de $p_i$ entre esa invocación y su respuesta correspondiente (si la hay), denotada $res(op)$, son pasos especificados por $A$ cuando $p_i$ invoca $op$.
\end{enumerate}

Se dice que una operación $op$ es completa si su invocación y su respuesta aparecen en la ejecución. Una operación está pendiente si en la ejecución solo aparece su invocación. Un proceso es correcto en una ejecución, si toma infinitos pasos. Por simplicidad, y sin pérdida de generalidad, identificamos la invocación de una operación con su primer paso y su respuesta con su último paso.\\

\subsection{Condiciones de progreso}

El comportamiento de los objetos concurrentes se describe por medio de propiedades de seguridad y viveza, a menudo denominadas condiciones de progreso \cite{ArtOfMpP}. La verificación de los algoritmos concurrentes se puede analizar a partir de dos aspectos fundamentales:
\begin{itemize}
    \item {Condición de progreso}: La cual garantiza que eventualmente algo bueno sucederá, también llamada la seguridad de los algoritmos.
    
    \item {Correctitud}: La cual garantiza que nada malo sucederá, también llamada viveza de los algoritmos
\end{itemize}

La correctitud la detallaremos en la siguiente sección. Por otro lado, existen diferentes especificaciones o condiciones de progreso, las cuales pueden ser bloqueantes o no bloqueantes. Las condiciones de progreso bloqueantes se puede llegar a producir un el retraso inesperado en cualquier proceso, esto puede retrasar la ejecución de otras instrucciones o el completo impedimento de que los demás procesos continúen. \\

\subsection{Condiciones de progreso bloqueantes}

Los métodos que tienen una condición de progreso bloqueante suelen estar basados en candados (en lenguaje Java son los {\it Lock}), cuyo funcionamiento se basa en hacer cumplir la exclusión mutua; más precisamente, los candados se usan para indicar que un hilo está ejecutando la sección crítica y estos {\it bloquean} la entrada a los demás hilos, de modo que se satisface la exclusión mutua. De tal forma que un candado es usado para aislar una sección crítica, de modo que cualquier hilo puede adquirir y liberar la sección crítica siempre y cuando otro hilo no se encuentre en ejecución. En lenguaje Java, cuando dos hilos ejecutan una sección crítica, uno la adquiere, y la bloquea, mientras que el otro hilo espera la respuesta de la solicitud para adquirir acceso a la sección crítica. Sin embargo, cabe mencionar que ante el fallo o suspensión del primer hilo, se producirá un bloqueo, ya que el segundo hilo podría nunca recibir respuesta del candado (Java posee infraestructura para capturar estas excepciones), pero cabe tenerlo en cuenta. Formalicemos un poco la exclusión mutua en los algoritmos.\\

Sea $SC_{\boldsymbol{A}}$ el intervalo durante el cual un hilo $\boldsymbol{A}$ ejecuta una sección crítica $SC$. Formalmente, para que un algoritmo satisfaga la exclusión mutua debe cumplir que las secciones críticas de diferentes hilos no se superponen, es decir, para los hilos $\boldsymbol{A}$ y $\boldsymbol{B}$, consideremos los intervalos $SC_{\boldsymbol{A}}$ y $SC_{\boldsymbol{B}}$ en los cuales se ejecutan las secciones críticas, entonces sucede $SC_{\boldsymbol{A}} \rightarrow SC_{\boldsymbol{B}}$ o $SC_{\boldsymbol{B}} \rightarrow SC_{\boldsymbol{A}}$. \\

Se puede mencionar dos condiciones de progreso bloqueantes, las cuales recurren al uso de candados: {Deadlock-Free}: Si algún hilo $\boldsymbol{A}$ intenta adquirir el candado, entonces existe algún hilo $\boldsymbol{B}$, puede ser el mismo hilo $\boldsymbol{A}$, que tendrá éxito en adquirir el candado (entrar a la sección crítica). {\it Starvation-free}: Todo hilo que intenta adquirir el candado tiene éxito en algún momento. También se puede decir que toda adquisición de candado eventualmente lo libera. \\

Nótese que la condición {\it Starvation-free} implica la condición {Deadlock-Free}. Además, la condición {Deadlock-Free} es importante, pues nos asegura que el sistema nunca se va a {\it bloquear}, ya que aunque los hilos individuales pueden atascarse, esperando la adquisición de un candado, para siempre (lo que se denomina inanición o {\it starvation}), pero entonces deben existir algunos hilos que sigan ejecutando la sección crítica.

\subsection{Condiciones de progreso no-bloqueantes}

Como hemos visto, las condiciones de progreso bloqueantes recurren al uso de candados para satisfacer la exclusión mutua, sin embargo, es posible definir otro tipo de condiciones de progreso, sin la necesidad de recurrir a los candados, estas son conocidas como condiciones de progreso no-bloqueantes, en las cuales el retraso inesperado de un proceso no retrasa a los demás procesos. \\

Dos condiciones no-bloqueantes muy importantes son las siguientes. La condición {\it wait-free}: Un método u operación termina de ejecutarse en número finito de pasos, se dice que un algoritmo es {\it wait-free} si todos los métodos u operaciones que lo componen lo son. 
La condición {\it non-blocking} o {\it lock-free}: Un método u operación es {\it non-blocking} si se garantiza que en una ejecución infinita de esta, existen infinitas operaciones que se completan en un número finito de pasos.\\

En otras palabras, la condición {\it wait-free} garantiza que todo proceso, siempre no haya fallado súbitamente, eventualmente progresa. Cualquier método {\it wait-free}  es también {\it non-blocking} pero no al revés. Los métodos {\it wait-free} pueden llegar a ser ineficientes, es por ello que una propiedad menos restrictiva como {\it non-blocking} es muy útil. Además, se podría decir que estas condiciones de progreso son la versión de las condiciones {\it Deadlock-Free} y {\it Starvation-free} que no recurren a la definición de candados. Esto es sumamente útil, pues en el modelo de computación concurrente que se presentó se consideran posibles fallas en los procesos. Aún más, el algoritmo que se muestra como principal contribución no recurre al uso de candados; sin embargo, más adelante se discute esto. 

\section{\label{sectionIII} Correctitud }

Diversas nociones de corrección para objetos concurrentes han sido propuestas; sin embargo, la mayoría se basan en alguna noción de equivalencia con el comportamiento secuencial. En \cite{ArtOfMpP} se presentan varias condiciones de corrección como la consistencia inactiva (Quiescent consistency), la consistencia secuencial (Sequential consistency) o la linealizabilidad. Una propiedad importante de la linealizabilidad es la de ser modular (también llamada local) \cite{Herlihy_1990}; esta misma propiedad hace de la linealizabilidad una condición de corrección más fuerte que las primeras dos. 

\subsection{Condición de correctitud: Linealizabilidad}

La linealizabilidad es la noción estándar utilizada para definir una implementación concurrente correcta de un objeto definido por una especificación secuencial. Intuitivamente, una ejecución es linealizable si las operaciones pueden ordenarse secuencialmente, sin reordenar operaciones que no se superpongan, de modo que sus respuestas satisfagan la especificación del objeto implementado.\\

Se define una especificación secuencial de un objeto concurrente $T$ es una máquina de estados especificada a través de una función de transición $\delta$. Dado un estado $q$ y una invocación $inv(op)$, $\delta (q, inv(op))$ devuelve la tupla $(q', res(op))$ (o un conjunto de tuplas si la máquina no es determinista) indicando que la máquina pasa al estado $q'$ y la respuesta a $op$ es $res(op)$. En nuestras especificaciones, $res(op)$ se escribe como un salto de tupla $\left< op\: :\:r \right>$, donde $r$ es el valor de salida de la operación. Las secuencias de tuplas invocación-respuesta, $ \left < inv(op) \: :\: res(op) \right > $, producidas por la máquina de estado, se denominan ejecuciones secuenciales.\\

Para formalizar la linealizabilidad debemos definir un orden parcial $<_{\alpha}$ en las operaciones completadas de una ejecución $\alpha$, por lo que denotamos $op <_{\alpha} op'$ si y solo si $res(op) \rightarrow inv(op')$ en $\alpha$. Decimos que dos operaciones son concurrentes si son incomparables por $<_{\alpha}$, usamos $op||_{\alpha} op'$ para denotar que dos operaciones son concurrentes.

\begin{definition}\label{LinealizabilidadDEF}
Linealizabilidad: Sea $A$ una implementación de un objeto concurrente $T$. Una ejecución $\alpha$ de $A$ es linealizable si existe una ejecución secuencial $S$ de $T$ tal que
\begin{itemize}
\item $S$ contiene todas las operaciones completadas de $\alpha$ y puede contener algunas operaciones pendientes. Las entradas y salidas de invocaciones y respuestas en $S$ concuerdan con las entradas y salidas en $\alpha$.
\item  Para cualesquiera operaciones completadas $op$ y $op'$ en $\alpha$, si sucede que $op <_{\alpha} op'$, entonces $op$ aparece antes que $op'$ en $S$.
\end{itemize}

Decimos que $A$ es linealizable si cada una de sus ejecuciones es linealizable.
\end{definition}

Decir que un objeto concurrente es correcto, lleva a tratar de encontrar una forma de extender el orden parcial a un orden total. La condición de corrección de linealizabilidad busca justamente esto. Como se ha mencionado anteriormente, dada una ejecución $\alpha$, la relación $<_{\alpha}$ es un orden parcial, cuando la ejecución es linealizable la relación $<_{\alpha}$ se convierte en un orden total. \\

Podemos explicar el concepto de linealizabilidad por medio de un ejemplo. Si consideramos, dos hilos A y B, cada vez que se ejecute el programa obtendremos una secuencia de llamadas y respuestas de operaciones. Denotemos por simplicidad la $i$-esíma llamadas en el hilo A (o B) como $IA-i$ (la cual hemos llamado anteriormente $inv(op)$), y la respuesta como $RA-i$ ($res(op)$). Claramente, se pueden dar superposiciones entre las llamadas a métodos, llamadas hechas por el hilo A o B. Pero cada evento (es decir, invocaciones y respuestas) tiene un orden en tiempo real. Entonces, las invocaciones y respuestas de todos los métodos llamados por A y B se pueden asignar a un orden secuencial, el orden puede como se muestra a continuación.

\begin{figure}[H]
    \centering
    \includegraphics[width=0.55\textwidth]{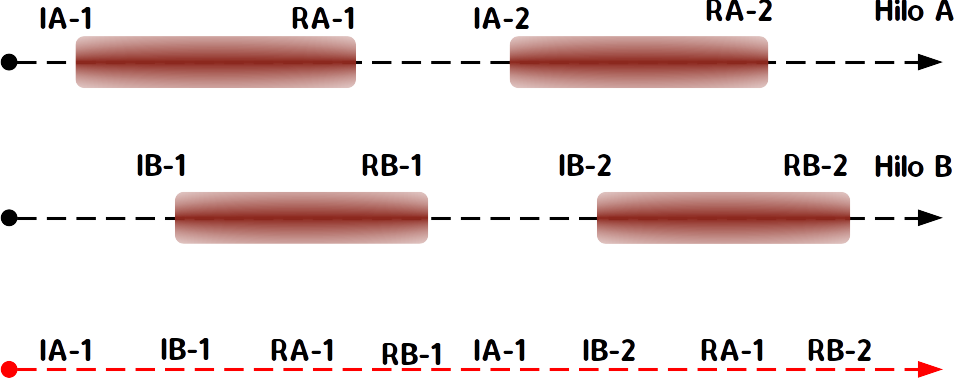}
    \caption{Secuencia de eventos en una ejecución en dos hilos A, B}
    \label{figF1}
\end{figure}

Una ordenación en tiempo real de los eventos se muestra en la flecha punteada roja. El orden de los eventos mostrado es:
$$
\mathrm{IA-1},\:\:\mathrm{IB-1},\:\:\mathrm{RA-1},\:\:\mathrm{RB-1},\:\:\mathrm{IA-2},\:\:\mathrm{IB-2},\:\:\mathrm{RA-2},\:\:\mathrm{RB-2},\:\:
$$
Esta secuencia de eventos representa una ejecución $\alpha$, en \cite{ArtOfMpP} se define de forma distinta y se le llama historial; sin embargo, son equivalentes. Los dos puntos de la definición de linealizabilidad informalmente significan que podemos dar una reordenación de los eventos siempre y cuando se respete el orden para las operaciones que cumplen $op<_{\alpha}op'$. Esto significa que, si el evento respuesta de una operación ocurrió antes que el evento de llamada de otra operación, entonces en la reordenación se debe preservar este orden. \\

Por ejemplo, las reordenaciones validas para la ejecución mostrada en la figura~\ref{figF1}, son:
\begin{enumerate}
    \centering
    \item  $\mathrm{IA-1},\:\: \mathrm{RA-1},\:\: \mathrm{IB-1},\:\: \mathrm{RB-1},\:\: \mathrm{IB-2},\:\: \mathrm{RB-2},\:\: \mathrm{IA-2},\:\: \mathrm{RA-2}$
    \item  $\mathrm{IB-1},\:\: \mathrm{RB-1},\:\: \mathrm{IA-1},\:\: \mathrm{RA-1},\:\: \mathrm{IB-2},\:\: \mathrm{RB-2},\:\: \mathrm{IA-2},\:\: \mathrm{RA-2}$
    \item  $\mathrm{IB-1},\:\: \mathrm{RB-1},\:\: \mathrm{IA-1},\:\: \mathrm{RA-1},\:\:\mathrm{IA-2},\:\: \mathrm{RA-2},\:\: \mathrm{IB-2},\:\: \mathrm{RB-2}$
    \item  $\mathrm{IA-1},\:\: \mathrm{RA-1},\:\: \mathrm{IB-1},\:\: \mathrm{RB-1},\:\:  \mathrm{IA-2},\:\: \mathrm{RA-2},\:\: \mathrm{IB-2},\:\: \mathrm{RB-2}$
\end{enumerate}
Formalmente, estas reordenaciones son ejecuciones secuenciales. Observemos que solo estamos permutando el orden de los eventos de llamada y respuesta de los métodos que son concurrentes. Debemos remarcar que esto es natural, ya que en métodos concurrentes no sabemos qué método se ejecuta primero, pues la duración de su ejecución es paralela. \\

\begin{figure}[H]
    \centering
    \includegraphics[width=0.55\textwidth]{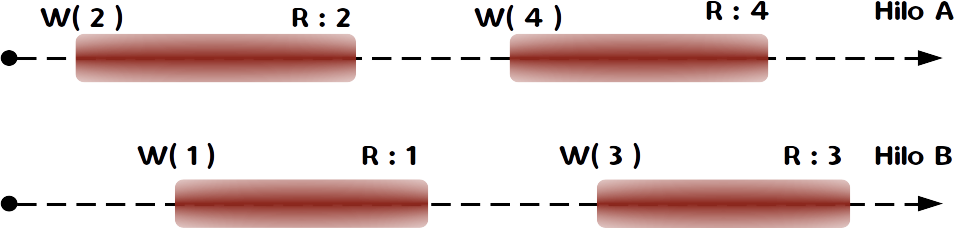}
    \caption{Ejecución de un método de lectura (W), escritura (R)}
    \label{figF2}
\end{figure}

Sin embargo, de estas ejecuciones secuenciales, ¿cómo podemos confirmar si nuestra ejecución es correcta? Aquí nos estamos refiriendo a la ejecución $\alpha$ de la definición~\ref{LinealizabilidadDEF}. La definición formal nos dice que, si al menos una ejecución secuencial es consistente con la ejecución $\alpha$, entonces la ejecución es linealizable. Podemos ver esto con un ejemplo sencillo. Considerando el ejemplo anterior, pensemos que las operaciones se refieren a invocaciones, un método que primero es escritura (denotada como evento $Writte: W$) y finaliza con lectura (evento $Read: R$), como que muestra en la siguiente figura~\ref{figF2}. De esta ejecución basta con tomar la ejecución secuencial 2), donde la secuencia es:
$$ Writte(1),\:\:Read(1),\:\:Writte(2),\:\:Read(2),\:\:Writte(3),\:\:Read(3),\:\:Writte(4),\:\:Read(4).$$

La secuencia es válida, ya que sigue la especificación secuencial de un registro, es decir, los {\it read} obtienen el valor del {\it write} más reciente.

\subsection{Condición de correctitud: Linealizabilidad por conjuntos}


La linealización de conjuntos nos permite linealizar varias operaciones en el mismo punto, es decir, todas estas operaciones se ejecutan simultáneamente. Se sabe que la linealización de conjuntos tiene estrictamente más poder de expresividad que la linealización; existe otro tipo denominado linealización de intervalos, la cual es estrictamente más poderosa que la linealización de conjuntos \cite{RelaxedQueuesAndStacks}. Además, la linealizabilidad y la linealización de conjuntos son propiedades composicionales\cite{Armando13}.\\

Una especificación conjunto-concurrente de un objeto concurrente difiere de una ejecución secuencial en que $\delta$ recibe como entrada el estado actual $q$ de la máquina y un conjunto $Inv= \{ inv(op_1),...,inv(op_t)  \}$ de invocaciones de operaciones, y $\delta(q,Inv)$ retorna $\delta(q',Res)$, donde $q'$ es el siguiente estado y $Res=\{  res(op_1),...,res(op_t) \}$ son las respuestas de las invocaciones en $Inv$. Los conjuntos $Inv$ y $Res$ son llamados clases de concurrencia. Debemos mencionar que una especificación de conjunto-concurrente donde las clases de concurrencia tenga un único elemento corresponde a una especificación secuencial.

\begin{definition}
Linealizabilidad por conjuntos: Sea $A$ una implementación de un objeto concurrente $T$. Una ejecución $\alpha$ de $A$ es linealizable por conjuntos si existe una ejecución de conjunto-concurrente $S$ de $T$ tal que
\begin{itemize}
\item $S$ contiene todas las operaciones completadas de $\alpha$ y puede contener algunas operaciones pendientes. Las entradas y salidas de invocaciones y respuestas en $S$ concuerdan con las entradas y salidas en $\alpha$.
\item  Para cualesquiera operaciones completadas $op$ y $op'$ en $\alpha$, si sucede que $op <_{\alpha} op'$, entonces $op$ aparece antes que $op'$ en $S$.
\end{itemize}

Decimos que $A$ es linealizable por conjuntos si cada una de sus ejecuciones es linealizable por conjuntos.
\end{definition}

Debemos resaltar que estos conjuntos son de hecho las clases de equivalencia de la relación $||_{\alpha}$, es decir, son las clases de equivalencia de las operaciones concurrentes. La interpretación es la misma que en la linealizabilidad, pero en este caso estamos considerando un conjunto de llamadas y un conjunto de respuestas.

\section{\label{sectionIV} Pilas Concurrentes por Conjuntos con Multiplicidad}
En términos generales, una pila con semántica relajada, que se denomina multiplicidad, permite a las operaciones simultáneas $Pop$ (quitar) obtengan el mismo elemento, pero todos los elementos se devuelven en orden LIFO y no se pierde ningún elemento. Formalmente, nuestra pila de conjuntos concurrentes se especifica de la siguiente manera:

\begin{definition}\label{defPilaSET}
El conjunto de elementos que se pueden poner ($Push$) es $\boldsymbol{N} = \{1, 2, . . .\}$, y el conjunto de estados $Q$ es el conjunto infinito de cadenas $N^{*}$. El estado inicial es la cadena vacía, indicada como $\boldsymbol{\epsilon}$. En el estado $\boldsymbol{q}$, el primer elemento en $\boldsymbol{q}$ representa la parte superior de la pila, que podría estar vacía si $\boldsymbol{q}$ es la cadena vacía. Las transiciones son las siguientes:
       
\begin{itemize}
\item Para $\boldsymbol{\hat{q}} \in Q:\:\:\:$
\begin{equation}\label{PilaMultiplicidad::Condition_1}
    \delta(\boldsymbol{\hat{q}}, \boldsymbol{Push(a)})=(
        \boldsymbol{\hat{q}*a},\left< \boldsymbol{Push(a)}:\boldsymbol{True}\right>
        )
\end{equation}

\item Para $\boldsymbol{\hat{q}*a} \in Q$, donde $\boldsymbol{a}\in \boldsymbol{N}$ y $t$ procesos:
\begin{equation}\label{PilaMultiplicidad::Condition_2}
    \delta(\boldsymbol{\hat{q}*a} ,\{\boldsymbol{Pop_1}(\:),...,\boldsymbol{Pop_t}(\:) \}  )= (\boldsymbol{\hat{q}},\{ \left<\boldsymbol{Pop_1}(\:): \boldsymbol{a} \right>,...,\left<\boldsymbol{Pop_t}(\:): \boldsymbol{a} \right>\})
\end{equation}

\item Para una pila vacía $\boldsymbol{\epsilon}\in Q$:
\begin{equation}\label{PilaMultiplicidad::Condition_3}
    \delta(\boldsymbol{\epsilon} , \boldsymbol{Pop}(\:) )= (\boldsymbol{\epsilon}, \left<\boldsymbol{Pop}(\:): \epsilon \right>)
\end{equation}
\end{itemize}
\end{definition}

Un lema de \cite{RelaxedQueuesAndStacks}, muestra que cualquier algoritmo que implemente la pila concurrente por conjuntos mantiene el comportamiento de una pila secuencial en ciertos casos. De hecho, la única razón por la que la implementación no proporciona capacidad de linealización se debe únicamente a las operaciones $Pop$ que son concurrentes. Aún más, una pila secuencial puede ser descrita por la definición~\ref{defPilaSET} si en el segundo punto se restringe a $t=1$ procesos.

\subsection{Algoritmo: SetStackLogic}
Con el fin de describir un algoritmo linealizable por conjuntos que sea una implementación de la pila con multiplicidad, primero se describe la estructura de los nodos que forman la pila. Es usual que las estructuras de datos, tipo cola o pila, estén formadas por listas de nodos, los cuales contienen elementos de tipos primitivos. Sin embargo, debido a la concurrencia y a la multiplicidad, es necesario definir una estructura más compleja en los nodos. \\

Además, para definir cierto tipo de operaciones es útil recurrir a los objetos de tipo \\ $\boldsymbol{AtomicReference}$, los cuales se proveen en Java. Las variables de tipo $\boldsymbol{AtomicReference}$ proporciona una referencia (a algún objeto) que se puede leer y escribir atómicamente. Es decir, que múltiples hilos que intentan modificar la misma referencia almacenada en la variable de tipo $\boldsymbol{AtomicReference}$ lo hacen de manera atómica, de modo que el estado de la referencia será consistente a las modificaciones. Supongamos por el momento que las referencias de una instancia dada de $\boldsymbol{AtomicReference}$ son objetos $\boldsymbol{nodo_i}$ (más adelante se definirá la estructura de los nodos), los métodos principales de cualquier instancia son:
\begin{itemize}
    \item $\boldsymbol{get(\:)}$: Este método permite que la referencia almacenada en la variable de tipo $\boldsymbol{AtomicReference}$ se pueda leer atómicamente.
    \item $\boldsymbol{CompareAndSet(nodo_1,nodo_2)}$: Este método comparar la referencia\\ almacenada en la instancia de $\boldsymbol{AtomicReference}$, llamémosle $\boldsymbol{nodo_0}$, contra una referencia esperada $\boldsymbol{nodo_1}$, y si las dos referencias son iguales (es decir $\boldsymbol{nodo_0 == nodo_1}$) entonces el método inserta la nueva referencia $\boldsymbol{nodo_2}$ en la instancia de $\boldsymbol{AtomicReference}$. Cuando el método $\boldsymbol{CompareAndSet}$ cambia la referencia en la instancia de $\boldsymbol{AtomicReference}$ este retorna $\boldsymbol{True}$ (regurlarmente nos referimos a este caso diciendo que el método $\boldsymbol{CompareAndSet}$ fue exitoso). En caso contrario, la referencia no sufre de ningún cambio y el método retorna $\boldsymbol{False}$ (nos referiremos a este caso diciendo que el método fue fallido o no exitoso).
\end{itemize}
\begin{algorithm}[H]
    \caption{Nodo(x)}\label{NodoStack}
    \begin{algorithmic}[1]
        \Statex $\boldsymbol{Nodo}$
        \State $\boldsymbol{value}:$  Entero de valor $\boldsymbol{x}$
        \State $\boldsymbol{next}:\:\boldsymbol{Nodo}$
        \State $\boldsymbol{elim}:\:\:$ Variable booleana 
    \end{algorithmic}
\end{algorithm}

Los nodos están definidos como la estructura que se muestra en el algoritmo~\ref{NodoStack}. Cada nodo contiene tres atributos, uno llamado $\boldsymbol{value}$, es cual es el valor de tipo primitivo almacenado en el nodo, supondremos que estos son enteros por simplicidad y concordancia con la definición \ref{defPilaSET}, el atributo $\boldsymbol{next}$ es un objeto de tipo $\boldsymbol{Nodo}$ cuya función es la de una referencia, y el atributo $\boldsymbol{elim}$ el cual es una variable booleana. El atributo $\boldsymbol{next}$ es una referencia al nodo siguiente de la lista de nodos. Por otro lado, cada nodo posee un atributo booleano $\boldsymbol{elim}$ cuya función es indicar si el nodo se encuentra o no en la pila; más adelante se formalizará esta noción, la cual se ha denominado {\it estado lógico}.

\begin{figure}[H]
    \centering
    \includegraphics[width=0.7\textwidth]{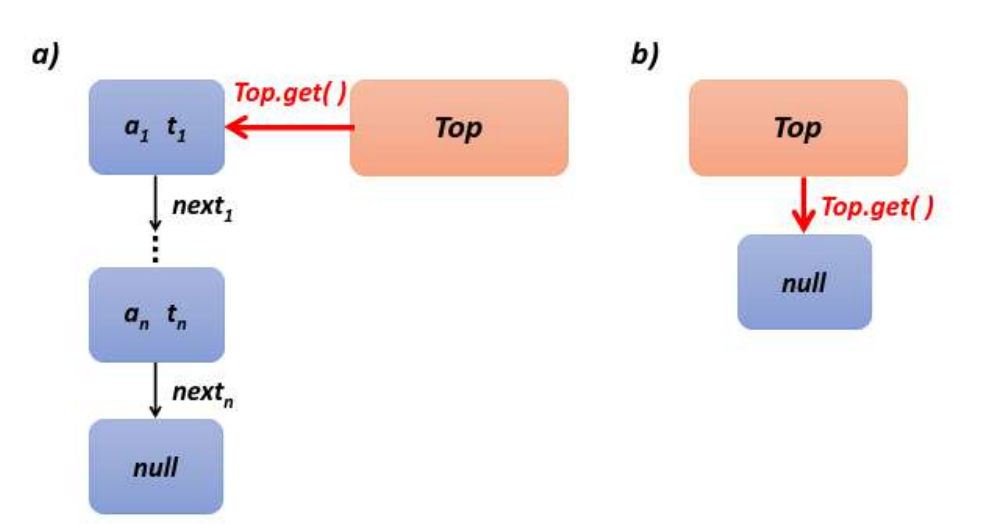}
    \caption{{\it Panel a)} Representación de una estructura de pilas, junto con la referencia atómica $\boldsymbol{Top}$. {\it Panel b)} Representación de una estructura de pilas sin elementos en la pila, es decir, la pila vacía $\boldsymbol{\epsilon}$.}
    \label{fig:pila__2}
\end{figure}

La estructura de pila se representa como una lista de nodos, véase figura \ref{fig:pila__2}. Es estas listas cada nodo posee una referencia, $\boldsymbol{next}$, la cual hace referencia al siguiente nodo de la lista, mientras que el valor de cada nodo es $\boldsymbol{value}$. La referencia del último nodo siempre apunta a un $\boldsymbol{null}$. Tengamos en cuenta el problema ABA al usar $\boldsymbol{CompareAnSet}$, para nuestro caso se puede evitar el problema ABA (y por claridad) suponiendo que cada nodo nuevo tiene una referencia distinta. Los nodos cuya variable booleana $\boldsymbol{elim}$ es $\boldsymbol{True}$ no son tomados en cuenta como parte de la pila, de modo que esta variable nos indica si el nodo forma parte de la pila. Además, el algoritmo consta de una variable de tipo $\boldsymbol{AtomicReference}<\boldsymbol{Nodo}>$:
\begin{itemize}
    \item $\boldsymbol{Top}:$ Contiene la referencia al primer nodo de la lista; $\boldsymbol{Top.get(\:)}$ es la referencia que apunta al primer nodo. Regularmente, llamaremos cabezal a la referencia atómica $\boldsymbol{Top}$, la cual se puede decir que encapsula a toda la pila.
\end{itemize}
En la figura \ref{fig:pila__2} {\it panel a)} se representa una pila del algoritmo SetStackLogic \ref{alg1-Pseudo}, en donde la referencia atómica $\boldsymbol{Top}$ apunta al primer nodo de la lista. Podemos decir que la lista de nodos está conformada por los pares de valores enteros y booleanos: $\left[\boldsymbol{(a_1,t_1)},\boldsymbol{(a_2,t_2)},...,\boldsymbol{(a_n,t_n)}\right]$, y los nodos que pertenecen a la pila son aquellos cuya variable booleana es $\boldsymbol{t_{i}}=\boldsymbol{False}$. En el {\it panel b)} se representa una pila vacía donde no hay ningún nodo y $\boldsymbol{Top}$ apunta a una referencia nula. Nótese que se puede representar perfectamente una pila vacía con una lista de nodos como en el {\it panel a)}, en donde todos los nodos posean $\boldsymbol{t_{i}}=\boldsymbol{True}$; el estado de la memoria de las pilas vacías no es único.\\

El algoritmo \ref{alg1-Pseudo}: {\it SetStackLogic}, que se presenta a continuación, está inspirado en el algoritmo {\it LockFreeStack} de \cite{ArtOfMpP}, el cual es un algoritmo linealizable de pilas {\it Lock Free}. 
\begin{algorithm}[H]
\caption{SetStackLogic}\label{alg1-Pseudo}
\begin{algorithmic}[1]
\Statex $\boldsymbol{Shared\:Variables}:$ Top
\Procedure{Push}{x}
\While{$True$}
\State $t$ = Top.get()
\If{t.elim == $False$}
    \State $x.$next = $t$
    \If{Top.CompareAndSet($t,x$)}
        \State \Return $True$
    \EndIf
\Else
    \State Top.CompareAndSet($t,t$.next)
\EndIf
\EndWhile
\EndProcedure 
\Statex
\Procedure{Pop}{}
\While{$True$}
\State $t$ = Top.get()
\If{$t=\epsilon$}
    \State \Return $\epsilon$
\EndIf
\If{$t$.elim == $false$}
    \State $t.$elim = $True$
    \State  Top.CompareAndSet($t,t.$next)
    \State \Return $t$.value
\Else
    \State Top.CompareAndSet($t,t$.next)
\EndIf
\EndWhile
\EndProcedure
\end{algorithmic}
\end{algorithm}
Después de discutir la estructura y representación  de las pilas, podemos pasar a describir las operaciones del algoritmo SetStackLogic. Las operaciones son las usuales (claro que ahora cumplirán multiplicidad): $\boldsymbol{Push}$ y $\boldsymbol{Pop}$.

\begin{itemize}
    \item $\boldsymbol{Push(x)}$:  Añade un nodo $\boldsymbol{x}$ en orden LIFO. La operación inicia un ciclo $\boldsymbol{While}$, en donde cada iteración es un intento para insertar el nodo $\boldsymbol{x}$ en la pila. En cada iteración se extrae el nodo $\boldsymbol{t}$ en la referencia del cabezal $\boldsymbol{Top}$. Si el nodo $\boldsymbol{t}$ ya fue lógicamente eliminado (línea 4, cuando $\boldsymbol{t.elim}==\boldsymbol{True}$), entonces avanza el cabezal $\boldsymbol{Top}$ al siguiente nodo por medio de un $\boldsymbol{CompareAndSet}$, en caso de que el nodo al que apunta el cabezal no haya sido lógicamente eliminado $\boldsymbol{t.elim}==\boldsymbol{False}$, conecta por medio de su referencia el nodo $\boldsymbol{t}$ al nodo $\boldsymbol{x}$. Después hace un cambio en el cabezal $\boldsymbol{Top}$ por medio del $\boldsymbol{CompareAndSet(t,x)}$, en caso de un exitoso el nodo $\boldsymbol{x}$ fue insertado correctamente y retorna en la línea 7, en caso contrario vuelve a iterar para otro intento.
    
    \item $\boldsymbol{Pop}$: Remueve un nodo en orden LIFO de la pila. La operación inicia un ciclo $\boldsymbol{While}$, en donde cada iteración es un intento para remover el primer nodo en la pila. En cada iteración se extrae el nodo $\boldsymbol{t}$ en la referencia del cabezal $\boldsymbol{Top}$. Si el nodo $\boldsymbol{t}$ es $\boldsymbol{\epsilon}$ la operación retorna un valor nulo, líneas 17 y 18, indicando que la pila está vacía. En caso contrario, cuando el nodo no es nulo, evalúa la línea 20, si $\boldsymbol{elim} \neq \boldsymbol{False}$, el nodo está lógicamente eliminado y pasa el cabezal al siguiente nodo en la línea 25 (Nótese que si la pila consta solamente de nodos eliminados lógicamente el cabezal avanza hasta llegar a la referencia nula). Cuando el nodo si está en la pila lógicamente $\boldsymbol{elim} == \boldsymbol{False}$ para a las líneas 21 a 23, primero elimina lógicamente el nodo en la línea 21 y después la referencia $\boldsymbol{Top}$ avanza al siguiente nodo en 22, finalmente retorna el valor del nodo eliminado $\boldsymbol{t.value}$ en la línea 23.
\end{itemize}
El siguiente teorema enuncia las propiedades que el algoritmo SetStackLogic cumple. Con esto se podrá apreciar y mostrar formalmente la noción del estado lógico de la pila y la correctitud del algoritmo.

\begin{theorem}\label{SetStackLogic}
El algoritmo SetStackLogic es una implementación linealizable por conjuntos de la pila con multiplicidad y es $non$-$blocking$.
\end{theorem}

La siguiente sección demuestra el teorema \ref{SetStackLogic}. La demostración esta divida de la siguiente forma, en la subsección \ref{StackLogicProof::1} se demuestra que el algoritmo \ref{alg1-Pseudo} satisface {\it non-blocking}, en la subsección \ref{Procedimientolinealización} se describe la forma en la que se construye la linealización por conjuntos, también en esta subsección se definen los {\it puntos de linealización}. En la subsección \ref{linealizaciónLogical} se muestra que usando los puntos de linealización de la subsección \ref{Procedimientolinealización} se obtiene efectivamente una linealización por conjuntos y en la subsección \ref{StackLogicProof::4} se demuestra que la linealización por conjuntos, dada por la subsección \ref{linealizaciónLogical}, es una implementación de la pila con multiplicidad, o más formalmente, se cumplen las transiciones de estado que corresponden a la pila, dada por la definición \ref{defPilaSET}. \\

Denotamos una operación $\boldsymbol{Pop}$ que devuelve el valor $\boldsymbol{x}$ por $\left<\boldsymbol{Pop}(\:): \boldsymbol{x} \right>$, mientras que una operación $\boldsymbol{Push}$ que inserta un valor $\boldsymbol{x}$ la denotamos por $\left<\boldsymbol{Push}(\boldsymbol{x}): True \right>$.

\section{\label{sectionV} Demostración formal}

\subsection{\label{StackLogicProof::1} Prueba de $non$-$blocking$: Set-Stack-Logic}

Veamos que el algoritmo es $non$-$blocking$. Supongamos una ejecución infinita del algoritmo Set-Stack-Logic, sea $\boldsymbol{op}$ una operación cualquiera.

\begin{itemize}
    \item Supongamos que $\boldsymbol{op}$ es una operación $\boldsymbol{Push}$, veamos que es {\it non-blocking}. Para que la operación no sea completada, las evaluaciones de las líneas 4 y 6 deben fallar una cantidad infinita de veces. Si la línea 4 falla, entonces debe existir una operación $\boldsymbol{Pop}$ la cual altero el valor de la variable $\boldsymbol{elim}$ del nodo $\boldsymbol{t}$, ejecutando la línea 21 (ya que por defecto todos los nodos poseen $\boldsymbol{t.elim}=\boldsymbol{False}$). Notemos que cualquier $\boldsymbol{Pop}$ que ejecuta la línea 21 retorna en dos instrucciones, y, por lo tanto, se completa. En resumen: Si la línea 4 falla, entonces debe existir una operación $\boldsymbol{Pop}$ completada. 
    
    Por otro lado, si la línea 6 falla siempre, sígnica que el método $\boldsymbol{compareAndSet}$ en la línea 6 falla en toda iteración, por lo que existe una operación $\boldsymbol{op'}$, ya sea $\boldsymbol{Pop}$ o $\boldsymbol{Push}$ la cual ejecute con éxito una operación $\boldsymbol{compareAndSet}$ para cada iteración. Veamos que debe existir una operación completada, si $\boldsymbol{op'}$ se completa es trivial. Supongamos que $\boldsymbol{op'}$ no está completada para toda iteración, entonces $\boldsymbol{compareAndSet}$ debió de ejecutarse en la línea 10, en caso de ser un $\boldsymbol{Push}$ o en la línea 25, en caso de ser un $\boldsymbol{Pop}$, sin embargo, dado que el estado inicial de la pila es $\boldsymbol{\epsilon}$ no pueden existir una cantidad infinita de nodos, alguna operación $\boldsymbol{Push}$ debe poner un nodo (lo cual implica que se completó) en algún momento para que la pila no se vacíe, y, por lo tanto, existe una operación completada siempre. 
    
    \item Para las operaciones $\boldsymbol{pop}$ es análogo, ya que si una operación $\boldsymbol{Pop}$ no se completa, entonces la línea 17 o 20 fallan infinitas veces. Si la línea 17 falla infinitas veces, el argumento es similar a lo visto, debe existir una operación $Push$ que ponga nodos. Para la línea 20 el argumento es el mismo que en el caso de la línea 4: Si la línea 20 falla, entonces debe existir una operación $\boldsymbol{Pop}$ completada. 
\end{itemize}

\subsection{\label{Procedimientolinealización}Procedimiento de linealización por Conjuntos: SetStackLogic}

Dado que el algoritmo es {\it non-blocking}, cualquier operación pendiente no bloquea al algoritmo, es decir, el algoritmo en si mismo se mantiene completando una infinidad de operaciones por cada operación que no se completa. Por lo que se puede suponer, sin perdida de generalidad, que cualquier ejecución finita $E$ no tiene operaciones pendientes, pues si las tuviera estas no bloquean a las demás. Sea $E$ una ejecución finita sin operaciones pendientes del algoritmo~\ref{alg1-Pseudo}: SetStackLogic.\\

En la linealización por conjuntos, el estado del objeto está codificado en la variable $\boldsymbol{Top}$ y en los valores de la etiqueta booleana $\boldsymbol{elim}$ de los nodos de la pila. Es decir, los elementos de la pila son los nodos cuyo valor booleano $\boldsymbol{elim}$ es $\boldsymbol{False}$, llamaremos a esto el estado lógico de la pila (más adelante se define formalmente).\\

Para linealizar la ejecución $E$, procederemos a construir una linealización $S$ de $E$ asignando un punto de linealización a cada operación $\boldsymbol{op}$ de $E$. El punto de linealización de $\boldsymbol{op}$, denotado $LinPt(\boldsymbol{op})$, es un evento primitivo $e$ entre la invocación $inv(\boldsymbol{op})$ y la respuesta $res(\boldsymbol{op})$.  Tengamos en cuenta que varias operaciones pueden tener el mismo punto de linealización. Por lo tanto, la linealización $S$ también se proporciona explícitamente. Finalmente, veremos que esta linealización $S$ induce una relación de orden total $<_{S}$ en las clases de concurrencia y que $S$ cumple con la especificación de pila~\eqref{defPilaSET}.   \\

Sea $\boldsymbol{op}$ en $E$, daremos las instrucciones de linealización definiendo primero el punto de linealización $LinPt(\boldsymbol{op})$, en el cual $\boldsymbol{op}$ pareciera tener efecto. Recordemos que  $LinPt(\boldsymbol{op})$ es un evento primitivo, lo cual significa que es la ejecución de alguna línea del algoritmo~\ref{alg1-Pseudo}, no una operación.

\begin{enumerate}
    \item Si $\boldsymbol{op}$ es una operación $\boldsymbol{Push}$ en $E$:
    Consideremos la última iteración de su ciclo $\boldsymbol{While}$, llamémosle $\boldsymbol{e_{CAS}}$ al paso que ejecuta la instrucción $\boldsymbol{CompareAndSet}$ de la línea 6 (este sería el único $\boldsymbol{CompareAndSet}$ exitoso). Linealizamos por conjuntos la operación $\boldsymbol{Push}$ en el punto $\boldsymbol{e_{CAS}}$, de modo es una clase de concurrencia por sí misma.
    
    \item Si $\boldsymbol{op}$ es una operación $\boldsymbol{Pop}$ en $E$, tenemos dos casos.
    \begin{itemize}
        \item[a)] Retorna la cadena vacía $\left<\boldsymbol{Pop}(\:): \boldsymbol{\epsilon} \right>$: Sea $\boldsymbol{op}$ cualquier operación $\boldsymbol{Pop}$ en $E$ tal que retorne en la línea 18 (Esto corresponde a las operaciones $\boldsymbol{Pop}$ que retornen la cadena vacía). Consideremos la última iteración en el ciclo $\boldsymbol{while}$, llamémosle $\boldsymbol{e_{get}}$ al paso correspondiente a la línea 16 de esta iteración, este paso es donde se extrae el nodo $\boldsymbol{t}$ almacenado de la variable $\boldsymbol{Top}$ por medio del método $\boldsymbol{get}()$. La condición en la línea $17$ es verdadera, pues estamos en la última iteración y la operación retorna en la línea 18, es decir, el nodo $\boldsymbol{t}$ tiene valor $\boldsymbol{\epsilon}$ en el paso $\boldsymbol{e_{get}}$. Linealizamos por conjuntos las operaciones $\boldsymbol{Pop}$ que retornan $\boldsymbol{\epsilon}$, en el punto $\boldsymbol{e_{get}}$, de modo es una clase de concurrencia por sí misma. \\
        
        \item[b)] Retorna un elemento no nulo $\left<\boldsymbol{Pop}(\:): \boldsymbol{x} \right>$: Sea $\boldsymbol{op}$ cualquier operación $\boldsymbol{Pop}$ en $E$ tal que retorne en la línea 23 (Esto corresponde a las operaciones $\boldsymbol{Pop}$ que retornen $\boldsymbol{t.value}$ tal que el valor de esta cadena es $\boldsymbol{x}$). \\
        
        Consideremos el conjunto $U_x$, tal que contiene todas las operaciones $\boldsymbol{Pop}$ en $E$ que devuelven el mismo elemento $\boldsymbol{x}$. Por suposición, cada elemento se coloca en la pila como máximo una sola vez, por lo tanto, cada operación $\boldsymbol{Pop} \in U_x$ extrae al mismo nodo en su última ejecución de la línea 16. Denotemos por $\boldsymbol{t_x}$ a este nodo, cuyo valor es $\boldsymbol{x}$. \\
        
        Observemos que todas las operaciones de $U_x$ extraen el nodo $\boldsymbol{t_x}$, lo cual corresponde a la ejecución de la línea 16. Consideremos el primer evento (de todas las operaciones $\boldsymbol{Pop}$ en $U_x$) que escribe en la variable booleana $\boldsymbol{elim}$ de $\boldsymbol{t}$ correspondiente a la línea 21, llamemos $\boldsymbol{e_{elim}^{x}}$ a este paso. Notemos que $\boldsymbol{e_{elim}^{x}}$ es el único paso en $U_x$ que realmente cambia el estado de la variable $\boldsymbol{elim}$, de modo que se elimina lógicamente el nodo $\boldsymbol{t_x}$ en el paso $\boldsymbol{e_{elim}^{x}}$. Linealizamos todas las operaciones en $U_x$ en el paso $\boldsymbol{e_{elim}^{x}}$, es decir, las operaciones en $U_x$ forman una clase de concurrencia ubicada en $\boldsymbol{e_{elim}^{x}}$.\\
    \end{itemize}
\end{enumerate}

Por simplicidad, llamémosle $\boldsymbol{op}$ a una clase de concurrencia de $S$. Observemos que podemos clasificar las clases de concurrencia de la siguiente manera:

\begin{enumerate}
    \item Se tiene $\boldsymbol{op}$ es una clase de concurrencia correspondiente a una operación $\boldsymbol{Push(a)}$ en $E$. 
    
    \item Se tiene $\boldsymbol{op}$ es una clase de concurrencia correspondiente a un conjunto operaciones $\boldsymbol{Pop(\:)}$ en $E$, tenemos dos subcasos:
    \begin{itemize}
        \item[a)] La clase de concurrencia $\boldsymbol{op}$ corresponde a una sola operación $\boldsymbol{Pop(\:)}$ tal que $\left<\boldsymbol{Pop}(\:): \boldsymbol{\epsilon} \right>$.\\
        
        \item[b)] La clase de concurrencia $\boldsymbol{op}$ corresponde a un conjunto de operaciones $U_x=\{\boldsymbol{Pop_1}(\:),...,\boldsymbol{Pop_t}(\:) \}$ tal que para toda operación $\boldsymbol{Pop_i(\:)}$ devuelve: $\left<\boldsymbol{Pop_i}(\:): \boldsymbol{x} \right>$ con $\boldsymbol{x}$ un elemento no nulo.\\
    \end{itemize}
\end{enumerate}

Estas clases de concurrencia que hemos definido dan la ejecución por conjuntos secuencial $S$. Observe que cada clase de concurrencia $\boldsymbol{op}$ de $S$ tiene un punto de linealización por construcción $LinPt(\boldsymbol{op})$, el cual es a su vez un punto de linealización para una o varias operaciones de $E$.\\

Ahora definamos $<_{S}$ para las clases de concurrencia, dadas dos clases de concurrencia $\boldsymbol{op},\boldsymbol{op'}$ en $S$, se define $\boldsymbol{op}<_{S} \boldsymbol{op'}$ si y solo si $LinPt(\boldsymbol{op}) \rightarrow LinPt(\boldsymbol{op'})$. Esta relación es de orden total sobre las clases de concurrencia de $S$.


\subsection{\label{linealizaciónLogical}linealización de la ejecución $E$}

Estas clases de concurrencia que hemos definido dan la ejecución secuencial por conjuntos $S$. Para mostrarlo primero veamos que dada una clase de concurrencia $\boldsymbol{op}$ de $S$, su punto de linealización $LinPt(\boldsymbol{op})$ siempre está entre los eventos de invocación y de respuesta de su conjunto de operaciones $\boldsymbol{op_E}$, es decir que cumple
\begin{equation}\label{Condicion1}
    inv(\boldsymbol{op_E}) \rightarrow  LinPt(\boldsymbol{op}) \rightarrow res(\boldsymbol{op_E}).
\end{equation}

Sea $\boldsymbol{op}$ en $S$, y $LinPt(\boldsymbol{op})$ su punto de linealización dado por las instrucciones en la subsección 3.2.1, probemos que se cumple~\eqref{Condicion1}.

\begin{enumerate}
    \item Si $\boldsymbol{op}$ es una clase de concurrencia correspondiente a una operación $\boldsymbol{Push(a)}$ de $E$. 
    
    Dado que $LinPt(\boldsymbol{op})=\boldsymbol{e_{CAS}}$, donde $\boldsymbol{e_{CAS}}$ es la ejecución de la instrucción $\boldsymbol{CompareAndSet}$ de la línea 6 de $\boldsymbol{Push(a)}$, de la última iteración de su ciclo $\boldsymbol{While}$. Claramente 
    $inv(\boldsymbol{Push(a)}) \rightarrow  \boldsymbol{e_{CAS}}$, pues la invocación precede a cualquier evento en las iteraciones del ciclo  $\boldsymbol{While}$, y $\boldsymbol{e_{CAS}} \rightarrow res(\boldsymbol{Push(a)})$, pues la respuesta de la operación $\boldsymbol{Push(a)}$ corresponde al evento asociado a la ejecución de la línea 7. 
    
    \item Si $\boldsymbol{op}$ es una clase de concurrencia correspondiente a un conjunto operaciones $\boldsymbol{Pop(\:)}$ en $E$, tenemos dos subcasos:
    \begin{itemize}
        \item[a)] La clase de concurrencia $\boldsymbol{op}$ corresponde a una sola operación $\boldsymbol{Pop(\:)}$ tal que $\left<\boldsymbol{Pop}(\:): \epsilon \right>$.\\

        El punto de linealización es $LinPt(\boldsymbol{op})=\boldsymbol{e_{get}}$, al igual que antes $inv(\boldsymbol{Pop(\:)}) \rightarrow  \boldsymbol{e_{get}}$, pues la invocación precede a cualquier evento en las iteraciones del ciclo  $\boldsymbol{While}$ y $\boldsymbol{e_{get}} \rightarrow res(\boldsymbol{Pop(\:)})$, pues la respuesta de la operación $\boldsymbol{Pop(\:)}$ corresponde al evento asociado a la ejecución de la línea 18. 
        
        \item[b)] La clase de concurrencia $\boldsymbol{op}$ corresponde a un conjunto de operaciones $U_x=\{\boldsymbol{Pop_1}(\:),...,\boldsymbol{Pop_t}(\:) \}$ tal que para toda operación $\boldsymbol{Pop_i(\:)}$ devuelve: $\left<\boldsymbol{Pop_i}(\:): \boldsymbol{x} \right>$ con $\boldsymbol{x}$ un elemento no nulo.\\
        
        En este caso, dado que estamos linealizando un conjunto de operaciones, debemos probar que, para toda operación $\boldsymbol{Pop_i}\in U_x$ se tiene 
        $$
        inv(\boldsymbol{Pop_i}) \rightarrow  LinPt(\boldsymbol{op}) \rightarrow res(\boldsymbol{Pop_i}),
        $$
        donde $U_x$ es el conjunto de todas las operaciones $\boldsymbol{Pop}$ en $E$ que devuelven el mismo elemento $\boldsymbol{x}$.
        
        Recordemos que $LinPt(\boldsymbol{op})=\boldsymbol{e_{elim}^{x}}$, con $\boldsymbol{e_{elim}^{x}}$ el evento descrito en la instrucción 2.b) del Procedimiento \ref{Procedimientolinealización}. Dado que $\boldsymbol{e_{elim}^{x}}$ es el primer evento que cambia el estado de $\boldsymbol{t.elim}$ de $\boldsymbol{False}$ a $\boldsymbol{True}$, debe pasar que $inv(\boldsymbol{Pop_i}) \rightarrow  \boldsymbol{e_{elim}^{x}}$, ya que en caso contrario $ \boldsymbol{e_{elim}^{x}} \rightarrow  inv(\boldsymbol{Pop_i})$, pero entonces la ejecución de la $\boldsymbol{t_x}==\boldsymbol{False}$ resultaría fallida lo que haría que pasara a la línea 25 alterando el valor de $\boldsymbol{Top}$ y en la siguiente iteración el nodo $\boldsymbol{t}$ tendría un valor distinto $\boldsymbol{y}\neq \boldsymbol{x}$, ya que el elemento $\boldsymbol{x}$ se inserta una única vez, contradiciendo el hecho de que $\boldsymbol{Pop_i}$ retorna $\boldsymbol{t_x}$. \\
        
        Por lo tanto, debe pasar que $inv(\boldsymbol{Pop_i}) \rightarrow  \boldsymbol{e_{elim}^{x}}$ para toda $\boldsymbol{Pop_i}$ en $U_x$. \\
        
        Para ver que $ \boldsymbol{e_{elim}^{x}} \rightarrow res(\boldsymbol{Pop_i}) $ es más fácil, ya que $\boldsymbol{e_{elim}^{x}}$ es el primer evento que cambia el estado de $\boldsymbol{t.elim}$, por lo que para cualquier $\boldsymbol{Pop_i}\in U_x$ se tiene que $\boldsymbol{e_{elim}^{x}} \rightarrow \boldsymbol{e_{elim}}$ donde $\boldsymbol{e_{elim}}$ es la ejecución de la línea 21 de $\boldsymbol{Pop_i}$ en la última iteración. Pero  $\boldsymbol{e_{elim}} \rightarrow res(\boldsymbol{Pop_i})$ y por transitividad de la precedencia $ \boldsymbol{e_{elim}^{x}} \rightarrow res(\boldsymbol{Pop_i})  $.
        
        Se concluye que: Para toda operación $\boldsymbol{Pop_i}$ en $U_x$ se tiene 
        $$
        inv(\boldsymbol{Pop_i}) \rightarrow  LinPt(\boldsymbol{op}) \rightarrow res(\boldsymbol{Pop_i}),
        $$
        donde $U_x$ es el conjunto de todas las operaciones $\boldsymbol{Pop}$ en $E$ que devuelven el mismo elemento $\boldsymbol{x}$.
        \end{itemize}
\end{enumerate}

Para concluir que $S$ una linealización por conjuntos de $E$ veamos lo siguiente. Por construcción, $E$ y $S$ tienen las mismas operaciones con las mismas respuestas. Considere las operaciones $\boldsymbol{op_1}$ y $\boldsymbol{op_2}$ de $E$ tal que $\boldsymbol{op_1}<_{E}\boldsymbol{op_2}$, por lo que $res(\boldsymbol{op_1})\rightarrow inv(\boldsymbol{op_2})$. En $S$, cada operación se linealiza en un paso que se encuentra entre la invocación y la respuesta de la operación. Por lo tanto, la clase de concurrencia de $\boldsymbol{op_1}$ aparece antes que la clase de concurrencia de $\boldsymbol{op_2}$ en $S$, pues se tiene 
$$
LinPt(\boldsymbol{op_1}) \rightarrow res(\boldsymbol{op_1})\rightarrow inv(\boldsymbol{op_2})  \rightarrow LinPt(\boldsymbol{op_2}) 
$$
por transitividad $LinPt(\boldsymbol{op_1}) \rightarrow LinPt(\boldsymbol{op_2})$, y por definición $\boldsymbol{\tilde{op_1}}<_{S}\boldsymbol{\tilde{op_2}}$, donde $\boldsymbol{\tilde{op_1}}$, $\boldsymbol{\tilde{op_1}}$ son las clases de concurrencia de $\boldsymbol{op_1}$ y $\boldsymbol{op_2}$ respectivamente. Concluimos que $S$ es una linealización por conjunto de $E$. Ahora falta ver que $S$ cumple con la especificación de pila.

\subsection{\label{StackLogicProof::4} Especificación de pila con multiplicidad: Algoritmo SetStackLogic}

En esta sección demostramos que el algoritmo SetStackLogic satisface la especificación de pila con multiplicidad. Para la demostración primero definiremos el estado lógico de la pila formalmente, lo cual nos facilitara la demostración:
\begin{definition}\label{EstadoLogico}
Sea $\boldsymbol{\hat{q}}\in \mathbb{N}^*$, el cual es de la forma
\begin{equation*}
    \boldsymbol{\hat{q}}= \boldsymbol{a_1}\boldsymbol{a_2}...\boldsymbol{a_k}
\end{equation*}
con $\boldsymbol{a_i}\in \mathbb{N} $ para $i\in \{1,...,k\}$.
Decimos que $\boldsymbol{\hat{q}}$ es el estado lógico que representa a la pila $\boldsymbol{t_1},...,\boldsymbol{t_m}$, si para cada elemento $\boldsymbol{a_i}$ existe un nodo $\boldsymbol{t_j}$ del algoritmo~\ref{alg1-Pseudo} tal que $\boldsymbol{t_j.elim}=\boldsymbol{False}$ y su valor es $\boldsymbol{a_i}$ con $\boldsymbol{t_1.next}=\boldsymbol{\epsilon}$ y  $\boldsymbol{t_1} = \boldsymbol{Top.get()}$
\end{definition}

Observemos que el estado de la memoria está representado por los nodos $\boldsymbol{\epsilon},\boldsymbol{t_1},...,\boldsymbol{t_m}$, mientras que el {\it estado lógico} esta representado por los nodos que están lógicamente en la pila; es decir, aquellos que cumplen $\boldsymbol{t_i.elim}=\boldsymbol{False}$, los nodos que no cumplen esta condición no participan en la pila de modo lógico. \\

{\it\bf SetStackLogic satisface la especificación de pila con multiplicidad. Demostración:}\\

Sea $E$ una ejecución finita sin operaciones pendientes del Algoritmo 1: SetStackLogic. Demostremos que cumple con la especificación de pila \ref{defPilaSET}. Donde el estado de la pila está dado por el estado lógico \ref{EstadoLogico}. \\

Sea $S$ la linealización por conjuntos de $E$ obtenida por el procedimiento de linealización \ref{Procedimientolinealización}. Tenemos que las clases de concurrencia están ordenadas como 
\begin{equation*}
    \boldsymbol{op_1}<_S\boldsymbol{op_2}<_S...<_S\boldsymbol{op_M}
\end{equation*}
Donde $op_i$ es la clase de concurrencia de alguna operación de $E$ y $M$ es el número de clases de concurrencia definidas por $<_S$. Sea $E_m$ el prefijo de $E$ con las operaciones $\{\boldsymbol{op_1}...\boldsymbol{op_m}\}$ y $S_m$ su correspondiente linealización por conjuntos, la cual de hecho induce el orden $\boldsymbol{op_1}<_S...<_S\boldsymbol{op_m}$.\\

Demostremos que $S_m$ satisface la especificación de pila \ref{defPilaSET} para toda $m\leq M$, cuyo estado está dado por $T_m$, por inducción sobre $m$. \\

El caso base es trivial, pues $E=\emptyset$, por lo que $S=\emptyset$, y el estado de la pila es el mismo después de ejecutar $S$ o $E$. \\

Como hipótesis de inducción supongamos que la ejecución secuencial por conjuntos $S_m$ de $E_m$ cumple con la especificación de pila, cuyo estado está dado por el estado lógico $T_m$. \\

Para el paso inductivo demostremos que se cumple para $E_{m+1}$ cuya linealización por conjuntos es $S_{m+1}$, dada por el procedimiento \ref{Procedimientolinealización}. Induce un orden total en las clases de equivalencia y podemos ordenarlas como:
\begin{equation*}
    \boldsymbol{op_0}<_{S}\boldsymbol{op_1}<_{S}...<_{S}\boldsymbol{op_{m+1}}
\end{equation*}
Por hipótesis de inducción el prefijo $E_m$ cumple con lo requerido, basta mostrar que al ejecutar la operación $\boldsymbol{op_{m+1}}$ la transición de estados corresponde a la especificación de pila \ref{defPilaSET}. \\

Sea $\boldsymbol{Top}$ la variable compartida antes de ejecutar $\boldsymbol{op_{m+1}}$, observemos que por hipótesis de inducción el estado de $\boldsymbol{Top}$ es el mismo que se obtiene de ejecutar $S_m$, por lo que se ejecutan las clases de concurrencia en orden: $\boldsymbol{op_0}<_{S}\boldsymbol{op_1}<_{S}...<_{S}\boldsymbol{op_m}$. \\

Sea $\boldsymbol{\hat{q}}\in \mathbb{N}^{*}$ el estado lógico de la pila después de ejecutar la clase de concurrencia $\boldsymbol{op_m}$, el cual representa el estado lógico de la pila $\boldsymbol{t_1},...,\boldsymbol{t_m}$.

\begin{enumerate}
    \item Si $\boldsymbol{op_{m+1}}$ es una clase de concurrencia correspondiente a una operación $\boldsymbol{Push(a)}$ en $E_m$:
    El punto de linealización de $\boldsymbol{op}$ es $LinPt(\boldsymbol{Push(a)})=\boldsymbol{e_{CAS}}$, el cual corresponde a la ejecución de la línea 6 con un $\boldsymbol{CompareAndSet}$ exitoso en la última iteración. Este punto también es el único evento en $\boldsymbol{Push}$ en el cual el estado lógico de la pila cambia, de $\boldsymbol{\hat{q}}$ a $\boldsymbol{\hat{q}*a}$. 
    
    Basta ver que si el estado en $\boldsymbol{op_m}$ es $\boldsymbol{\hat{q}}$, con $\boldsymbol{\hat{q}}\in \mathbb{N}^*$, el estado después de ejecutar $\boldsymbol{Push(a)}$ es $\boldsymbol{\hat{q}*a}$. De este modo se cumpliría formalmente la transición:
    \begin{equation*}
        \delta(\boldsymbol{\hat{q}}, \boldsymbol{Push(a)})=(
        \boldsymbol{\hat{q}*a},\left< \boldsymbol{Push(a)}:\boldsymbol{True}\right>
        ).
    \end{equation*}
    
    Dado que $\boldsymbol{op_m}<_{S} \boldsymbol{op_{m+1}}$, podemos asegurar que el estado de la pila en el evento $\boldsymbol{e_{CAS}}$ cambian de $\boldsymbol{\hat{q}}$ a $\boldsymbol{\hat{q}*a}$, pues si él estado de la pila es diferente de $\boldsymbol{\hat{q}}$, entonces algún evento lo modifico entre $\boldsymbol{op_m}$ y $\boldsymbol{op_{m+1}}$. 
        
    Podemos asegurar que no existe ningún evento que modifique el estado lógico de la pila $\boldsymbol{\hat{q}}$, pues los únicos eventos que cambian el estado lógico son las ejecuciones de la línea 6 o 21 (en caso de un $\boldsymbol{CompareAndSet}$ exitoso, el cual añada un nodo, para la línea 6) pero justo estos eventos corresponderían a puntos de linealización de alguna operación $\boldsymbol{op'}$, de modo que $\boldsymbol{op_{m}}<_{S}\boldsymbol{op'}<_{S} \boldsymbol{op_{m+1}}$, contradiciendo que el orden que induce la linealización por conjuntos $S_{m+1}$.
    
    Por lo tanto, el estado lógico después del evento $LinPt(\boldsymbol{op_{m}})$ y antes del evento $\boldsymbol{e_{CAS}}$ es $\boldsymbol{\hat{q}}$, pero en $\boldsymbol{e_{CAS}}$ el método $\boldsymbol{CompareAndSet}$ es exitoso, cambiando el estado lógico de la pila a $\boldsymbol{\hat{q}*a}$, ya que inserta el nodo $\boldsymbol{t}$ con valor $\boldsymbol{a}$ tal que $\boldsymbol{t.elim}=\boldsymbol{False}$. Ahora el estado en memoria es $\boldsymbol{t_1},...,\boldsymbol{t_m},\boldsymbol{t}$ cuyo estado lógico es $\boldsymbol{\hat{q}*a}$, pues el orden lo fija la lista $\boldsymbol{t_1},...,\boldsymbol{t_m},\boldsymbol{t}$. Entonces, para la operación $\boldsymbol{op_{m+1}}$ se cumple la transición:
    \begin{equation*}
        \delta(\boldsymbol{\hat{q}}, \boldsymbol{Push(a)})=(
        \boldsymbol{\hat{q}*a},\left< \boldsymbol{Push(a)}:\boldsymbol{True}\right>
        )
    \end{equation*}
    Por lo que concluimos que $S_{m+1}$ produce una ejecución secuencial por conjuntos de pila válida que es consistente con~\eqref{defPilaSET}, donde estado de la pila está representado por el estado lógico codificado en $\boldsymbol{\hat{q}*a}$, representado en la variable $\boldsymbol{Top}$

    
    \item Si $\boldsymbol{op_{m+1}}$ es una clase de concurrencia correspondiente a un conjunto operaciones $\boldsymbol{Pop(\:)}$ en $E_m$, tenemos dos casos:
    \begin{itemize}
        \item[a)] La clase de concurrencia $\boldsymbol{op_{m+1}}$ corresponde a una sola operación $\boldsymbol{Pop(\:)}$ tal que $\left<\boldsymbol{Pop}(\:): \epsilon \right>$.\\
        
        El punto de linealización es $\boldsymbol{e_{get}}$, correspondiente a la línea 16 de la última iteración. 
        Además, la operación $\boldsymbol{Pop(\:)}$ retorna en la línea 18. Por suposición
        \begin{equation*}
            LinPt(\boldsymbol{op_m}) \rightarrow \boldsymbol{e_{get}}.
        \end{equation*}
        
        Por lo que el estado lógico de la pila después de ejecutar $LinPt(\boldsymbol{op_m})$ debe de ser $\boldsymbol{\epsilon}$, pues en caso contrario en el evento  $\boldsymbol{e_{get}}$ se obtendría un nodo $\boldsymbol{t}\neq \boldsymbol{\epsilon}$, y no se ejecutaría la línea 4 satisfactoriamente, contradiciendo $\left<\boldsymbol{Pop}(\:): \epsilon \right>$.  
        
        Además, observemos que las operaciones $\boldsymbol{Pop}$ que devuelven una cadena nula: $\left<\boldsymbol{Pop}(\:): \boldsymbol{\epsilon} \right>$, en realidad no alteran el estado lógico de la pila, pues nunca ejecutan la línea 21. Por lo que se ejecute o no $\boldsymbol{op}$ el estado lógico permanece invariante, y es $\boldsymbol{\epsilon}$. Dicho en otras palabras, para la operación $\boldsymbol{op_{m+1}}$ se cumple la transición:
        \begin{equation*}
             \delta(\boldsymbol{\epsilon} , \boldsymbol{Pop}(\:) )= (\boldsymbol{\epsilon}, \left<\boldsymbol{Pop}(\:): \boldsymbol{\epsilon} \right>).
        \end{equation*}
        
        Concluimos que $S_{m+1}$ produce una ejecución secuencial por conjuntos de pila válida que es consistente con~\eqref{defPilaSET}, donde estado de la pila está representado por el estado lógico codificado en $\boldsymbol{\epsilon}$, representado en la variable $\boldsymbol{Top}$

        \item[b)] La clase de concurrencia $\boldsymbol{op_{m+1}}$ corresponde a un conjunto de operaciones $U_x=\{\boldsymbol{Pop_1}(\:),...,\boldsymbol{Pop_t}(\:) \}$ tal que para toda operación $\boldsymbol{Pop_i(\:)}$ devuelve: $\left<\boldsymbol{Pop_i}(\:): \boldsymbol{x} \right>$ con $\boldsymbol{x}$ un elemento no nulo. Recordemos que el conjunto $U_x$ es tal que contiene todas las operaciones $\boldsymbol{Pop}$ en $E_{m+1}$ que devuelven el mismo elemento $\boldsymbol{x}$. \\
        
        El punto de linealización es $LinPt(\boldsymbol{op})=\boldsymbol{e_{elim}^{x}}$, el cual corresponde a la primera ejecución de la línea 21 de todas las operaciones $\boldsymbol{Pop_x}$ en $U_x$, este punto también es el único evento en el cual el estado lógico de la pila cambia (cambiando $\boldsymbol{t_x.elim}=\boldsymbol{False}$ a $\boldsymbol{t_x.elim}=\boldsymbol{True}$ en el evento $\boldsymbol{e_{elim}^{x}}$ para $\boldsymbol{t_x}$ el nodo que contiene el valor de $\boldsymbol{x}$). \\
        
        Basta ver que si el estado en $\boldsymbol{op_{m}}$ es $\boldsymbol{\hat{q}}=\boldsymbol{\hat{p}*a}$, con $\boldsymbol{\hat{p}}\in \mathbb{N}^*$ y $\boldsymbol{a}\in \mathbb{N}$, el estado después de ejecutar $\boldsymbol{op_{m+1}}$ es $\boldsymbol{\hat{p}}$. Dado que $\boldsymbol{op_m}<_{S} \boldsymbol{op_{m+1}}$, podemos asegurar que el estado de la pila en el evento $\boldsymbol{e_{elim}^{x}}$ cambian de $\boldsymbol{\hat{p}*a}$ a $\boldsymbol{\hat{p}}$, pues si el estado la pila cambia entre $\boldsymbol{op_m}$ y $\boldsymbol{op_{m+1}}$, entonces algún evento lo cambio. Podemos asegurar que no existe ningún evento que cambie el estado lógico de la pila $\boldsymbol{\hat{p}*a}$, pues los únicos eventos que cambian el estado lógico son las ejecuciones de la línea 6 o 21 (en caso de un $\boldsymbol{CompareAndSet}$ exitoso, el cual añada un nodo, para la línea 6) pero justo estos eventos corresponderían a puntos de linealización de alguna operación cuya, con clase de concurrencia $\boldsymbol{op'}$, de modo que $\boldsymbol{op_m}<_{S}\boldsymbol{op'}<_{S} \boldsymbol{op_{m+1}}$, contradiciendo el orden dado por $S_{m+1}$. \\
        
        Por lo tanto, antes del evento $\boldsymbol{e_{elim}^{x}}$ el estado lógico es $\boldsymbol{\hat{p}*a}$. Además, podemos asegurar que para el nodo $\boldsymbol{t_m}=\boldsymbol{Top.get()}$ se tiene que $\boldsymbol{t_m.elim}==\boldsymbol{False}$, por la linea 20. Entonces, dado que es el ultimo nodo, por definición, su valor es $\boldsymbol{a}$. Entonces, $\boldsymbol{e_{elim}^{x}}$ cambia $\boldsymbol{t_m.elim}=\boldsymbol{True}$ con $\boldsymbol{t_m}$, cambiando el estado lógico de la pila a $\boldsymbol{\hat{p}}$, pues ahora no esite nodo alguno cuyo valor sea $\boldsymbol{a}$ y este logicamente en la pila. También se tiene que $\boldsymbol{x}=\boldsymbol{a}$, pues la línea 23 devuelve $\boldsymbol{t_a.value}$ el cual es $\boldsymbol{a}$. Dicho en otras palabras, para la operación $\boldsymbol{op_{m+1}}$ se cumple la transición:
        \begin{equation*}
             \delta(\boldsymbol{\hat{p}*a} ,\{\boldsymbol{Pop_1}(\:),...,\boldsymbol{Pop_t}(\:) \}  )= (\boldsymbol{\hat{p}},\{ \left<\boldsymbol{Pop_1}(\:): \boldsymbol{a} \right>,...,\left<\boldsymbol{Pop_t}(\:): \boldsymbol{a} \right>\}).
        \end{equation*}
        
        Por lo que concluimos que $S_{m+1}$ produce una ejecución secuencial de pila válida que es consistente con~\eqref{defPilaSET}, donde estado de la pila está representado por el estado lógico codificado en $\boldsymbol{\hat{q}}$, representado en la variable $\boldsymbol{Top}$.
    \end{itemize}
\end{enumerate}
Por lo que concluimos que $S_{m}$ produce una ejecución secuencial de pila válida que es consistente con~\eqref{defPilaSET} para toda $m\leq n$. Por consiguiente es válido para $E$ y $S$.


\section{\label{sectionVI} Funcionamiento}

Esta sección está destinada para explicar el funcionamiento del algoritmo SetStackLogic bajo algunos posibles escenarios, así como mostrar la transición de estados de forma explicita. Principalmente, para dar ejemplos de como se puede llegar a comportar la memoria del sistema durante una ejecución concurrente. Cada hilo se representará por medio de una línea negra punteada, los eventos se representan como $\boldsymbol{e}$ más algunos índices.\\

\begin{figure}[H]
    \centering
    \includegraphics[width=0.8\textwidth]{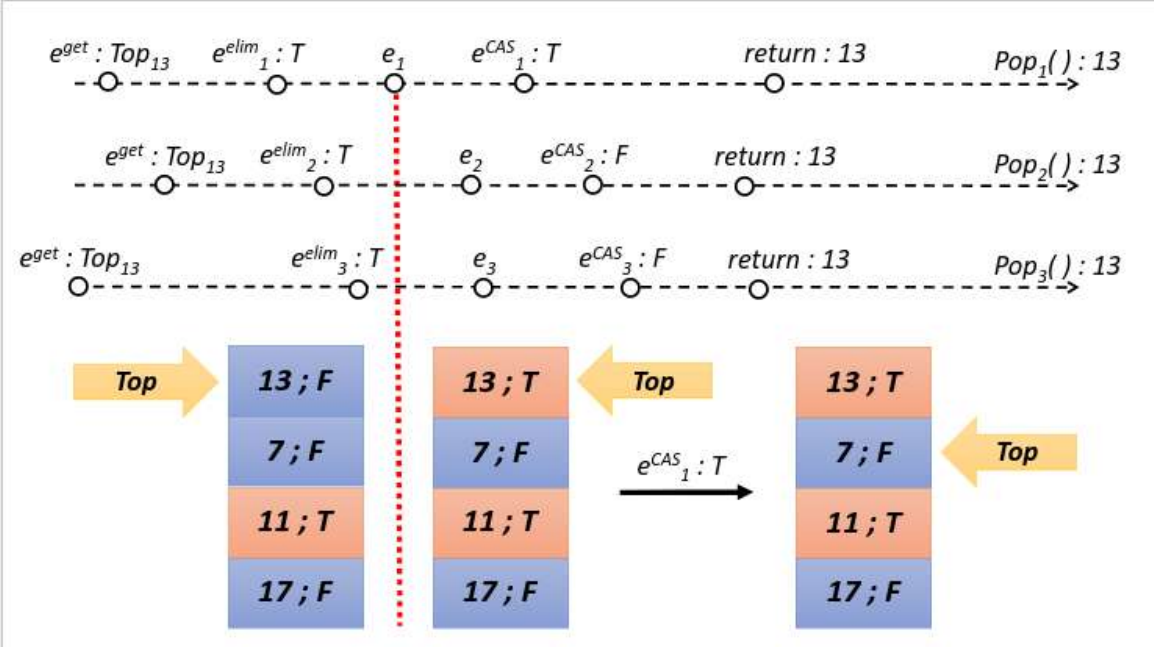}
    \caption{Representación de un conjunto dado de operaciones $\boldsymbol{Pop_i}$ que retornan todas el mismo valor. En este caso el conjunto es la clase de concurrencia de las operaciones.}
    \label{fig:pilasSecD__1}
\end{figure}

Para los siguientes dos ejemplos consideremos tres hilos que se ejecutan concurrentemente. El primer ejemplo está representado en la figura \ref{fig:pilasSecD__1}, donde cada hilo ejecuta una operación $\boldsymbol{Pop_i}$, además consideraremos el caso donde todas las operaciones retornan el mismo elemento. Como se puede apreciar durante las demostraciones, la iteración más relevante en las operaciones es siempre la última, en este ejemplo consideremos que estamos en la última iteración de cada $\boldsymbol{Pop_i}$. Consideremos que las operaciones ocurren de la siguiente forma:

\begin{itemize}
    \item La pila está constituida por la lista de nodos $$\left[\boldsymbol{(17,False)},\boldsymbol{(11, True)}, \boldsymbol{(7,False)}, \boldsymbol{(13,False)}\right].$$ El estado lógico de la pila es $\boldsymbol{(17,7,13)}$, siendo el nodo con valor 13 el primero en la pila.
    
    \item La instrucción de la línea 3 se representa por medio de $\boldsymbol{e^{get}}$, donde cada operación extra al nodo de la referencia por medio de $\boldsymbol{Top.get(\:)}$. Dado que en esta iteración todas las operaciones $\boldsymbol{Pop_i}$ finalizaron en valor de dicho nodo es 13, y su valor booleano $\boldsymbol{False}$.

    \item Todas las operaciones encuentran un nodo distinto de $\boldsymbol{\epsilon}$, línea 17 a 19.

    \item Cada operación ejecuta la línea 20 (evento representado por $\boldsymbol{e^{elim}_{i}}$) y dado que el primer nodo es $\boldsymbol{(13,False)}$ todas las operaciones pasan a la línea 21.

    \item Llamémosle $\boldsymbol{e_{i}}$ a la ejecución de la línea 21. Consideremos, sin perdida de generalidad, que la operación $\boldsymbol{Pop_1}$ es la primera en ejecutar dicha línea. Por lo tanto, $\boldsymbol{e_1}$ es el punto de linealización de este conjunto de operaciones y en este evento es donde se produce un cambio en el estado lógico de la pila, esto se representa por la línea punteada roja. 

    \item Después del evento $\boldsymbol{e_1}$ el estado lógico de la pila cambio a $\boldsymbol{(17,7)}$, pero veamos que la referencia $\boldsymbol{Top}$ aún apunta al nodo $\boldsymbol{(13,True)}$. Todas las operaciones ejecutan ahora la línea 22 (evento $\boldsymbol{e^{CAS}_i}$), el primer $\boldsymbol{CompareAndSet}$ es el que cambia la referencia de $\boldsymbol{Top}$ (no tiene mucha relevancia cuál es el primero). Finalmente, todas las operaciones retornan 13.

    \item En este último evento $\boldsymbol{e^{CAS}_i}$ la referencia atómica $\boldsymbol{Top}$ cambia, de modo que la memoria real de la pila cambia tal como se representa por medio de la flecha negra en la figura \ref{fig:pilasSecD__1}; sin embargo, el verdadero cambio del estado lógico se encuentra en la línea roja puntead, el cual es también el punto de linealización por conjuntos.
\end{itemize}

\begin{figure}[H]
    \centering
    \includegraphics[width=0.8\textwidth]{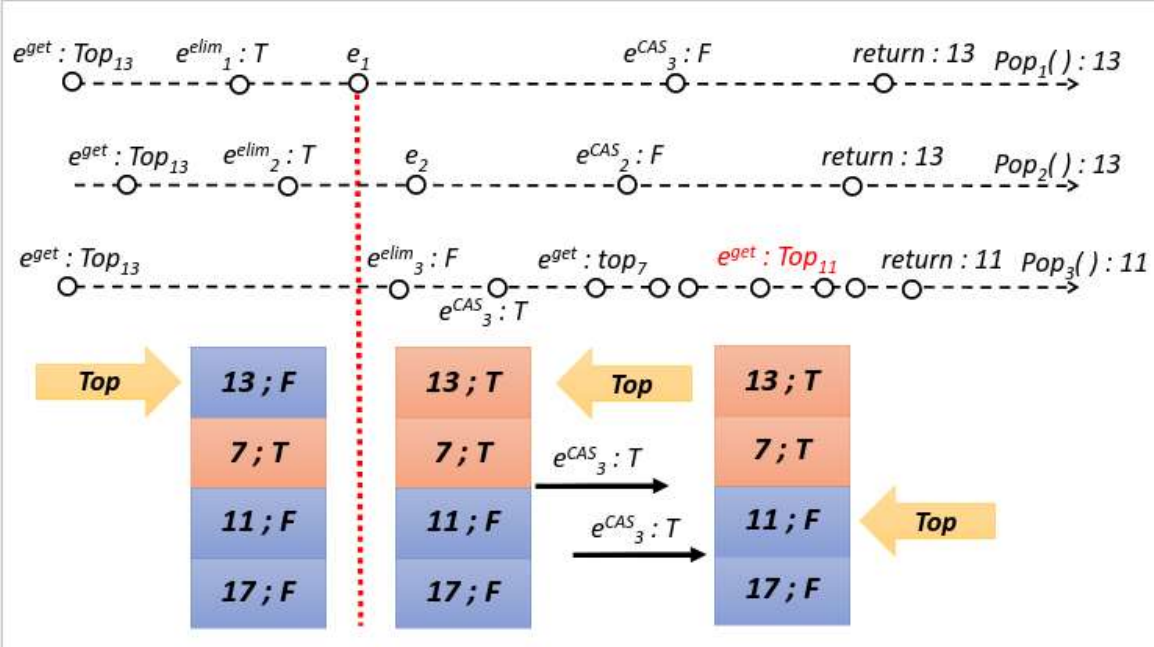}
    \caption{Representación de un conjunto dado de operaciones $\boldsymbol{Pop_i}$ que retornan diferentes valores. En este caso se tendrían dos clases de concurrencia.}
    \label{fig:pilasSecD__2}
\end{figure}

Como siguiente ejemplo consideremos un conjunto de operaciones $\boldsymbol{Pop}_i$, representado en la figura \ref{fig:pilasSecD__2}, las operaciones ocurren de la siguiente manera:

\begin{itemize}
    \item La pila está constituida por la lista de nodos $$\left[\boldsymbol{(17,False)},\boldsymbol{(11, False)}, \boldsymbol{(7,True)}, \boldsymbol{(13,False)}\right].$$ El estado lógico de la pila es $\boldsymbol{(17,11,13)}$.
    
    \item La instrucción de la línea 3 se representa por medio de $\boldsymbol{e^{get}}$. Todas las operaciones extraen la misma referencia al nodo $\boldsymbol{(13,False)}$.

    \item Usando la misma notación donde, la línea 20 es el evento representado por $\boldsymbol{e^{elim}_{i}}$ y $\boldsymbol{e_{i}}$ a la ejecución de la línea 21. En este caso los eventos $\boldsymbol{e^{elim}_i}$ las operaciones $\boldsymbol{Pop_1}$ y $\boldsymbol{Pop_2}$ ocurren antes del evento $\boldsymbol{e_1}$, en donde el estado lógico de la pila cambia. Pero el evento $\boldsymbol{e^{elim}_{3}}$ ocurre después, de modo que el $\boldsymbol{if}$ no permite ejecutar la línea 21 para $\boldsymbol{Pop_3}$.
    
    \item Las operaciones $\boldsymbol{Pop_1}$ y $\boldsymbol{Pop_2}$ se completan de una forma similar al caso anterior. En el evento $\boldsymbol{e^{CAS}_{3}}$ la referencia $\boldsymbol{Top}$ cambia de apuntar al nodo $\boldsymbol{(13,True)}$ al nodo $\boldsymbol{(7,True)}$ (instrucción de la línea 25). Observemos que el nodo $\boldsymbol{(7,True)}$ no está lógicamente en la pila.

    \item La operación $\boldsymbol{Pop_3}$ vuelve a iterar para otro intento de remover el nodo al inicio de la pila. Dado que el nodo $\boldsymbol{(7,True)}$ no está lógicamente en la instrucción 20, falla y pasa otra vez a la línea 25, donde se realiza otro $\boldsymbol{e^{CAS}_{3}}$ para cambiar la referencia de $\boldsymbol{Top}$ del nodo $\boldsymbol{(7,True)}$ al nodo $\boldsymbol{(11,False)}$.
    
    \item La operación $\boldsymbol{Pop_3}$ vuelve a iterar, extrayendo el nodo $\boldsymbol{e^{get}}$ del cabezal y después se ejecutan las líneas 20 a 23 retornando el valor 11.
    
\end{itemize}

\begin{figure}[H]
    \centering
    \includegraphics[width=0.8\textwidth]{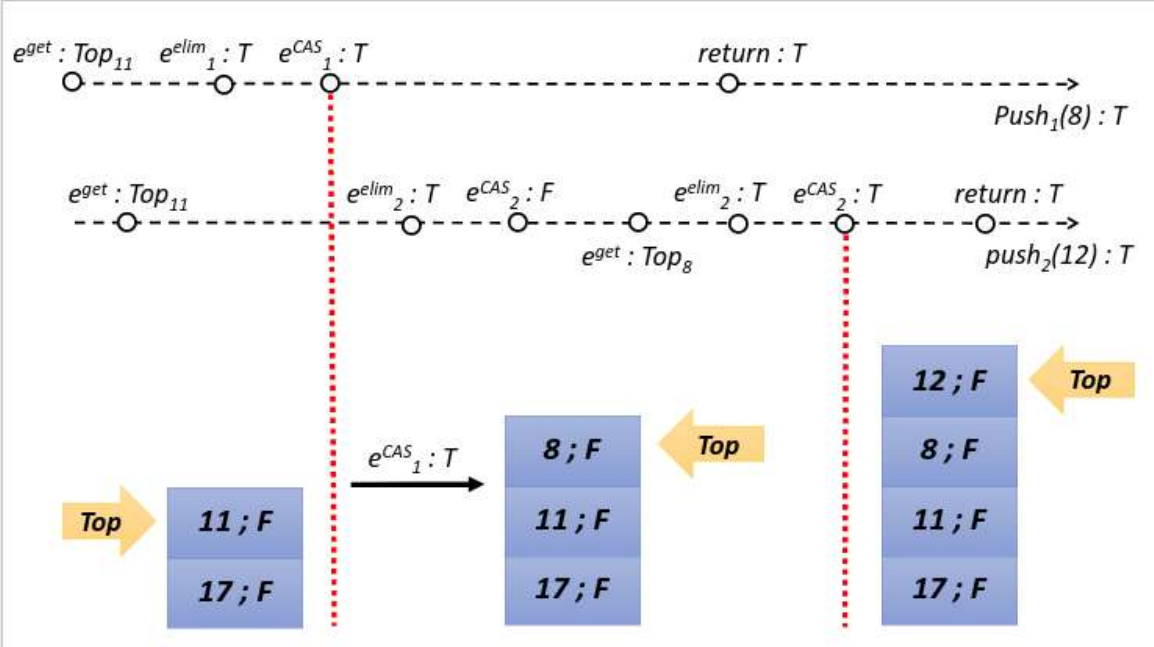}
    \caption{Representación de la interacción entre dos operaciones concurrentes: $\boldsymbol{Push_1((8,False)}$ y $\boldsymbol{Push_2((12,False))}$.}
    \label{fig:pilasSecD__3}
\end{figure}

Consideremos ahora un conjunto de operaciones $\boldsymbol{Push_1((8,False)}$ y $\boldsymbol{Push_2((12,False))}$, representado en la figura \ref{fig:pilasSecD__3} ,por simplicidad ejemplificaremos solo un par de operaciones, las cuales ocurren de la siguiente manera:

\begin{itemize}
    \item La pila está constituida por la lista de nodos $$\left[\boldsymbol{(17,False)},\boldsymbol{(11, False)}\right].$$ El estado lógico de la pila es $\boldsymbol{(17,11)}$.
    
    \item La instrucción de la línea 3 se representa por medio de $\boldsymbol{e^{get}}$. Ambas operaciones extraen la misma referencia al nodo $\boldsymbol{(11,False)}$.

    \item En la línea 5 la operación $\boldsymbol{Push_1}$ conecta el nuevo nodo $\boldsymbol{(8,False)}$ al primero nodo, en el evento y después en la línea 6 se realiza un $\boldsymbol{CompareAndSet}$ el cual cambia la referencia hacia el nuevo nodo, esto en el evento $\boldsymbol{e^{CAS}_{1}}$. En este último evento el estado lógico y la memoria real cambian, tan como se muestra en la figura. Después de este cambio de estado lógico, la operación $\boldsymbol{Push_2}$ ejecuta un $\boldsymbol{CompareAndSet}$ en el evento $\boldsymbol{e^{CAS}_{2}}$; sin embargo, falla, pues la referencia ya cambio.

    \item La operación $\boldsymbol{Push_2}$ vuelve a iterar siguiente un comportamiento similar al de la operación $\boldsymbol{Push_1}$, insertando el nuevo nodo $\boldsymbol{(12,False)}$ en el evento $\boldsymbol{e^{CAS}_{2}}$. 

    \item El estado final de la pila después de la ejecución de las operaciones $\boldsymbol{Push_1}$ y $\boldsymbol{Push_2}$ es
    $$\left[\boldsymbol{(17,False)},\boldsymbol{(11, False)}, \boldsymbol{(8,False)}, \boldsymbol{(12,False)}\right].$$
    Observemos que en este caso existen dos puntos de linealización distintos, cada uno en la última iteración de su respectiva operación $\boldsymbol{Push_i}$, la cual es una clase de concurrencia por sí misma.
\end{itemize}

Observemos que una vez que ocurre el punto de linealización de la primera operación, la segunda operación inevitablemente fallara en el $\boldsymbol{CompareAndSet}$; es decir, en el evento $\boldsymbol{e_{2}^{CAS}}$ se retorna $\boldsymbol{False}$. Esto es debido a que cambia el estado lógico de la pila en el punto de linealizacion. Esta propiedad nos permite linealizar por conjunto las operaciones $\boldsymbol{Push}$ en clases de concurrencia de un único elemento. Mientras que las operaciones $\boldsymbol{Pop}$ poseen un punto de linealización que permite que un conjunto dado de operaciones puedan ser linealizadas en un mismo punto de linelizacion.
\begin{figure}[H]
    \centering
    \includegraphics[width=0.8\textwidth]{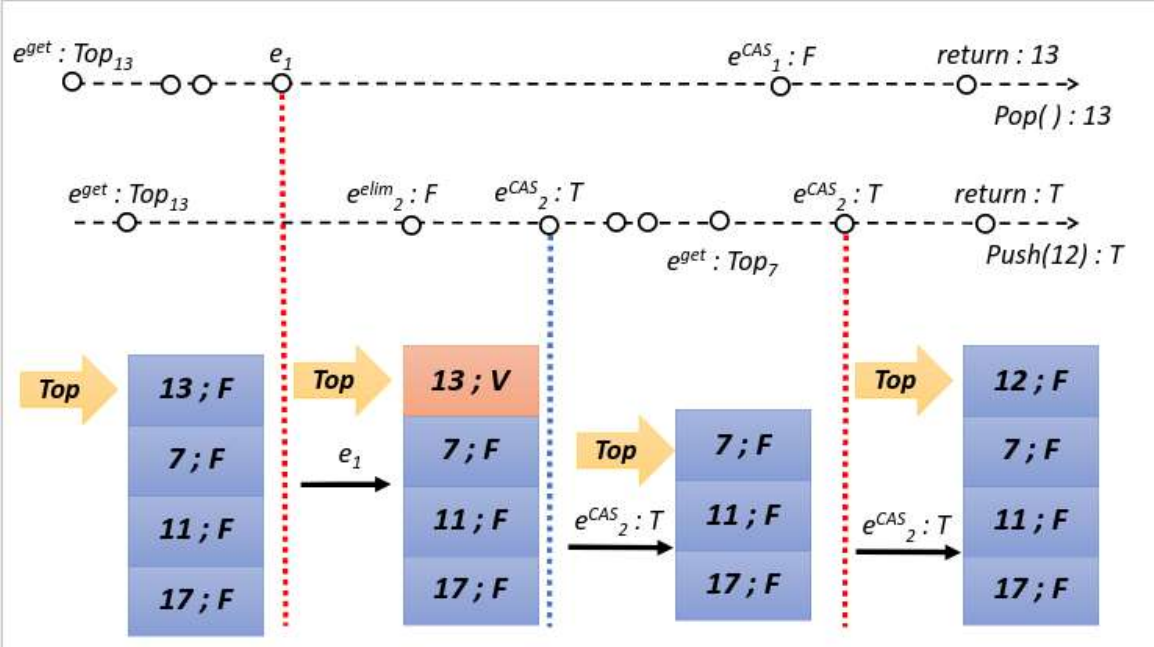}
    \caption{Representación de la interacción entre dos operaciones concurrentes: $\boldsymbol{Push((12,False))}$ y $\boldsymbol{Pop(\:)}$.}
    \label{fig:pilasSecD__4}
\end{figure}

Como último ejemplo consideremos una operación $\boldsymbol{Push((12,False))}$ y otra $\boldsymbol{Pop}$, representado en la figura~\ref{fig:pilasSecD__4}, con el siguiente comportamiento:

\begin{itemize}
    \item La pila está constituida por la lista de nodos $$\left[\boldsymbol{(17,False)},\boldsymbol{(11, False)},\boldsymbol{(7, False)},\boldsymbol{(13, False)}\right].$$ El estado lógico de la pila es $\boldsymbol{(17,11,7,13)}$.
    
    \item La instrucción de la línea 3 y 16 se representa por medio de $\boldsymbol{e^{get}}$. Ambas operaciones extraen la misma referencia al nodo $\boldsymbol{(13,False)}$.

    \item La operación $\boldsymbol{Pop}$ ejecuta primero la instrucción de la línea 21, en donde ocurre el punto de linealización $\boldsymbol{e_1}$. La pila cambia su estado lógico como se muestra en la primera transición, pero la referencia continúa apuntando al nodo $\boldsymbol{(13,True)}$.

    \item La operación $\boldsymbol{Push((12,False))}$ ejecuta la línea 4 (fallando, pues $\boldsymbol{(13,True).elim}==\boldsymbol{True}$), y después la línea 10, donde hace un cambio de referencia de $\boldsymbol{Top}$ al nodo $\boldsymbol{(7,False)}$.

    \item La ejecución del $\boldsymbol{CompareAndSet}$ en la línea 22 falla, independientemente de esto la operación se completa retornando el valor del nodo en la línea 23.

    \item En la operación $\boldsymbol{Push}$ inicia otra iteración y obtiene de $\boldsymbol{Top}$ la referencia al nodo $\boldsymbol{(7,False)}$. Después de ejecutar los debidos eventos llega a la línea 6 y en el evento $\boldsymbol{e^{CAS}_2}$ ejecuta un $\boldsymbol{CompareAndSet}$ exitoso, por lo que cambia la referencia al nuevo nodo añadido a la pila, el estado de la pila cambia en $\boldsymbol{e^{CAS}_2}$ y termina en $\boldsymbol{(17,11,7,12)}$.
\end{itemize}

Observemos que no hay problema si en lugar de una operación $\boldsymbol{Pop}$ consideramos un conjunto de operaciones: ya sea de 3 operaciones, como en el ejemplo de la figura \ref{fig:pilasSecD__1}, o un conjunto arbitrario de operaciones pero todas pertenecientes a la misma clase de concurrencia. Esto es gracias a que la única operación que cambia el estado lógico en la pila es aquella que contiene el punto de linealización del conjunto de operaciones $\boldsymbol{Pop_i}$ (siempre y cuando retornen el mismo elemento).

\section*{Agradecimientos}

Este trabajo fue financiado por el programa de Becas Nacionales para Estudio de Posgrado 2021 de CONACyT. Se agradecen los comentarios y discusiones al Dr. Armando Castañeda Rojano, quien brindo de sus conocimientos y experiencia que ayudaron en gran medida a la realización de este trabajo, el cual es resultado parcial de Tesis de Maestría.

\bibliographystyle{apsrev4-2}
\bibliography{main}
\end{document}